\documentclass[useAMS,usenatbib]{mn2e}
\usepackage{epsfig, color}

\title[Pop III Rotation and Structure]{Rotation and Internal Structure of Population~III Protostars}

\author[A. Stacy, T. H. Greif, R. S. Klessen, V. Bromm and A. Loeb]
       {Athena Stacy$^{1}$\thanks{E-mail: athena.stacy@nasa.gov}, Thomas H. Greif$^{2,3}$, Ralf S. Klessen$^{4}$, Volker Bromm$^{5}$, and Abraham Loeb$^{3}$\\
        $^{1}$NASA Goddard Space Flight Center, Greenbelt, MD 20771, USA \\
        $^{2}$Max-Planck-Institut f\"{u}r Astrophysik, Karl-Schwarzschild-Str. 1, 85741 Garching, Germany \\
        $^{3}$Astronomy Department, Harvard University, 60 Garden Street, Cambridge, MA 02138, USA \\
        $^{4}$Universit\"{a}t Heidelberg, Zentrum f\"{u}r Astronomie, Institut f\"{u}r Theoretische Astrophysik, Albert-Ueberle-Str. 2, 69120 Heidelberg, Germany\\
        $^{5}$Department of Astronomy and Texas Cosmology Center, University of Texas, Austin, TX 78712, USA}

\pagerange{\pageref{firstpage}--\pageref{lastpage}}

\pubyear{2012}

\begin{document}

\maketitle

\topmargin-1cm

\label{firstpage}

\begin{abstract}
We analyze the cosmological simulations performed in the recent work of \cite{greifetal2012}, which followed the early growth and merger history of Pop III stars while resolving scales as small as 0.05 R$_{\odot}$.  
This is the first set of cosmological simulations to self-consistently resolve the rotation and internal structure of Pop III protostars.
We find that Pop III stars form under significant rotational support which is maintained for the duration of the simulations.  
The protostellar surfaces spin from
 $\sim$50\% to nearly 100\% of Keplerian rotational velocity.
 
These rotation rates persist after experiencing multiple stellar merger events.
In the brief time period simulated ($\sim$ 10 yr), the protostars show little indication of convective instability, and their properties furthermore show little correlation with the properties of their host minihaloes.  
If Pop III protostars within this range of environments generally form with high degrees of rotational support, and if this rotational support is maintained for a sufficient amount of time, this has a number of crucial implications for Pop~III evolution and nucleosynthesis, as well as
the possibility for Pop III pair-instability supernovae, and
the question of whether the first stars produced gamma-ray
bursts.
\end{abstract}


\section{Introduction}

The formation of the first stars marked a crucial transition in the evolution of the early Universe (\citealt{bromm&larson2004,loeb2010}).   These stars were metal-free, and are thus also known as Population III (Pop III).  They are believed to have formed at $z \ga 20$ within dark matter minihaloes of mass $\sim$ 10$^6$ M$_{\odot}$  (e.g. \citealt{haimanetal1996,tegmarketal1997,yahs2003}).  
Pop III stars that were sufficiently massive then formed the first H{\sc ii} regions, beginning the process of reionization 
(e.g. \citealt{kitayamaetal2004,syahs2004,whalenetal2004,alvarezetal2006,johnsongreif&bromm2007}).
Stars within certain mass ranges  ended their lives as supernovae, thereby contributing to the initial enrichment of the intergalactic medium (IGM) with heavy
elements (\citealt{madauferrara&rees2001,moriferrara&madau2002,brommyoshida&hernquist2003,wada&venkatesan2003,normanetal2004,tfs2007,greifetal2007,greifetal2010,wise&abel2008,maioetal2011}; recently reviewed in 
 \citealt{karlssonetal2012}).  In particular, 
 non-rotating
 primordial stars with mass 140~M$_{\odot}$~$<$~$M_{*}$~$<$~260~M$_{\odot}$ are believed to have exploded as pair-instability supernovae (PISNe; \citealt{heger&woosley2002}), 
 releasing the entirety of their metal content into the IGM and surrounding haloes, while  stars within the range  15~M$_{\odot}$~$<$~$M_{*}$~$<$~40 M$_{\odot}$ ended their lives as core-collapse SNe.  
 On the other hand,  
 non-rotating
 Pop III stars with main sequence masses in the range 40~M$_{\odot}$~$<$~$M_{*}$~$<$~140~M$_{\odot}$ or  $M_{*}$~$>$~260~M$_{\odot}$ are expected to collapse directly into black holes, 
 thus contributing no metals to their surroundings.

The typical mass of Pop III stars is therefore a crucial factor in determining their role in the evolution of the early Universe.  However, the rotation of a star can also strongly influence stellar evolution and death,
(e.g. \citealt{maeder1987}; see also reviews by \citealt{maeder&meynet2000,maeder&meynet2012,langer2012}),  with stronger effects for lower-metallicity stars (e.g. \citealt{brottetal2011}).  For instance, instead of a normal core-collapse SN, rapidly rotating stars may end their lives as extremely energetic hypernovae (e.g. \citealt{nomotoetal2003}).  A further consequence may be that rapid rotation lowers the minimum mass at which Pop III stars can become PISNe  (\citealt{chatz&wheeler2012,yoonetal2012}),  possibly to masses as low as 65 M$_{\odot}$.  This occurs through the process of rotationally induced mixing, which in turn leads to chemically homogeneous evolution (CHE) and a larger final oxygen core mass as compared with a non-rotating star of the same initial mass.

Theoretical studies of stellar evolution furthermore find that, compared to non-rotating stars of the same mass, 
the typical effective temperatures and luminosities of rotating Pop~III stars are larger, and for an extended range of masses CHE shifts stellar evolutionary tracks blueward (\citealt{yoonetal2012}). This depends sensitively, however, on how angular momentum redistribution by magnetic fields is modeled.  Earlier work which does not implement the Spruit-Taylor dynamo (\citealt{spruit2002}), for instance, find that Pop III stars will not undergo CHE and will in fact end their evolution in a redder part of the Hertzsprung-Russell diagram (Ekstr{\"o}m et al. 2008).
The metal production within rotating primordial stars is generally higher, and rapidly-rotating models can lead to $^{14}$N yields that are several orders of magnitude larger than for corresponding non-rotating or slowly-rotating models (\nocite{ekstrometal2008}Ekstr{\"o}m et al. 2008; \citealt{yoonetal2012}).  However, we also note that 
these authors as well as
\cite{heger&woosley2010} 
somtimes
find significant $^{14}$N yields even in their non-rotating models
due to chemical mixing through convection between the helium and hydrogen-burning shells. 

Rotation may additionally determine a Pop~III star's potential for producing a gamma-ray burst (GRB), particularly given the connection between long-duration GRBs and the death of massive stars (e.g. \citealt{woosley&bloom2006}). This would provide a promising method of directly probing Pop III stars in their final stages, provided that Pop III GRBs occurred with sufficient frequency 
(\citealt{bromm&loeb2002,bromm&loeb2006,gouetal2004,belczynskietal2007,naoz&bromberg2007}). The collapsar model of GRB formation requires the presence of sufficient angular momentum in the progenitor for an accretion torus to form around the remnant black hole (e.g. \citealt{woosley1993,lee&ramirez2006}). Furthermore, to enable the escape of the accompanying relativistic jet from the star, the progenitor must also lose its hydrogen envelope (e.g. \citealt{zhangetal2004}; but see \citealt{suwa&ioka2011}). However, this latter requirement poses a difficulty for single-star progenitors, because removal of the hydrogen envelope will also lead to a decrease of angular momentum in the core (e.g. \citealt{spruit2002,hegeretal2005,petrovicetal2005}).  On the other hand, both conditions for collapsar GRB formation may be met in the case of
 close binary systems.  
As discussed in, e.g., \cite{macfadyen&woosley1999}, a binary companion can serve to remove the H envelope of the massive star, which then becomes a helium star.
Subsequently, the spin-orbit tidal interactions will spin up the He star (e.g. \citealt{leeetal2002,izzardetal2004}).  
However, the numerical calculations of \cite{detmersetal2008} question how often this scenario leads to a collapsar GRB.  They find that the tidal interactions between a Wolf-Rayet (WR) star and a compact object in a binary system will more often lead to a merger event instead of a collapsing and rapidly rotating star.


Another route to GRB formation arises for Pop III stars with sufficient spin, due to the effects of rapid rotation on a star's nucleosynthesis as well as its evolution off of the main sequence (MS; e.g. 
\citealt{yoon&langer2005,woosley&heger2006}; \citealt{yoonetal2012}).  For instance, \cite{yoon&langer2005} and \cite{woosley&heger2006} find that low-metallicity massive stars ($\ga$ 20 M$_{\odot}$) with rotation rates above 
$\sim$ 40-50\% of their breakup speeds  may undergo rotationally induced mixing and CHE, allowing the star to bypass the red giant phase and become a WR star.  This may allow the star to retain enough angular momentum to become a GRB, particularly if the star also undergoes little mass loss through stellar winds, as is expected for low and zero-metallicity stars 
(\citealt{kudritzki2002, vink&dekoter2005}).  
\cite{yoonetal2012} find that this route to GRB production through CHE applies to metal-free stars as well.


 \begin{figure*}
\includegraphics[width=.8\textwidth]{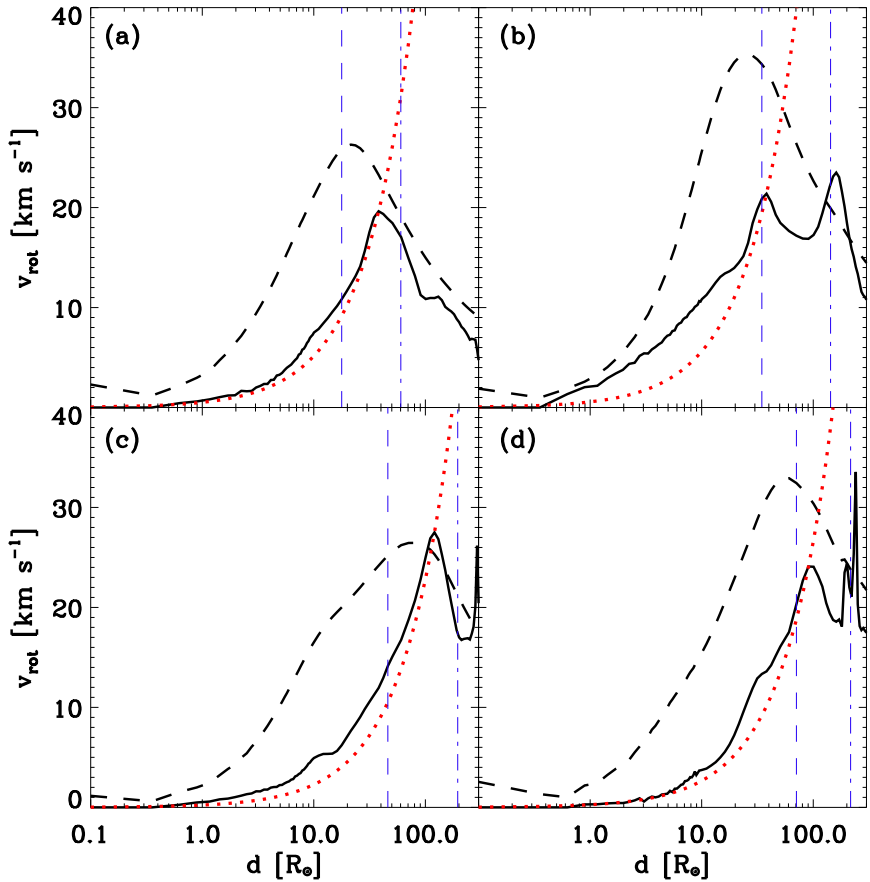}
 \caption{Profile of rotational velocity within the most massive protostar of each minihalo, shown with the solid lines and measured from the centers of the protostars.  
 {\it (a):}  Minihalo 1.
{\it (b):}  Minihalo 2.
{\it (c):}  Minihalo 3.
{\it (d):} Minihalo 4.
 Dashed lines show $v_{\rm Kep}$ for comparison.  Vertical blue dash-dot lines denote the photospheric surface of the protostar, $R_{\rm p}$.  
 Vertical blue dashed lines show location of the hydrostatic surface of the star, $R_*$.
 Dotted red lines show example profiles of solid-body rotation.  Note the significant rotational support within large portions of the protostars.  
The secondary peaks beyond the protostellar edge in panels {\it (b)}, {\it (c)}, and {\it (d)} are due to smaller protostars that are in tightly bound orbits around the main protostars.  
 }
\label{vrot}
\end{figure*}

 \begin{figure*}
\includegraphics[width=.8\textwidth]{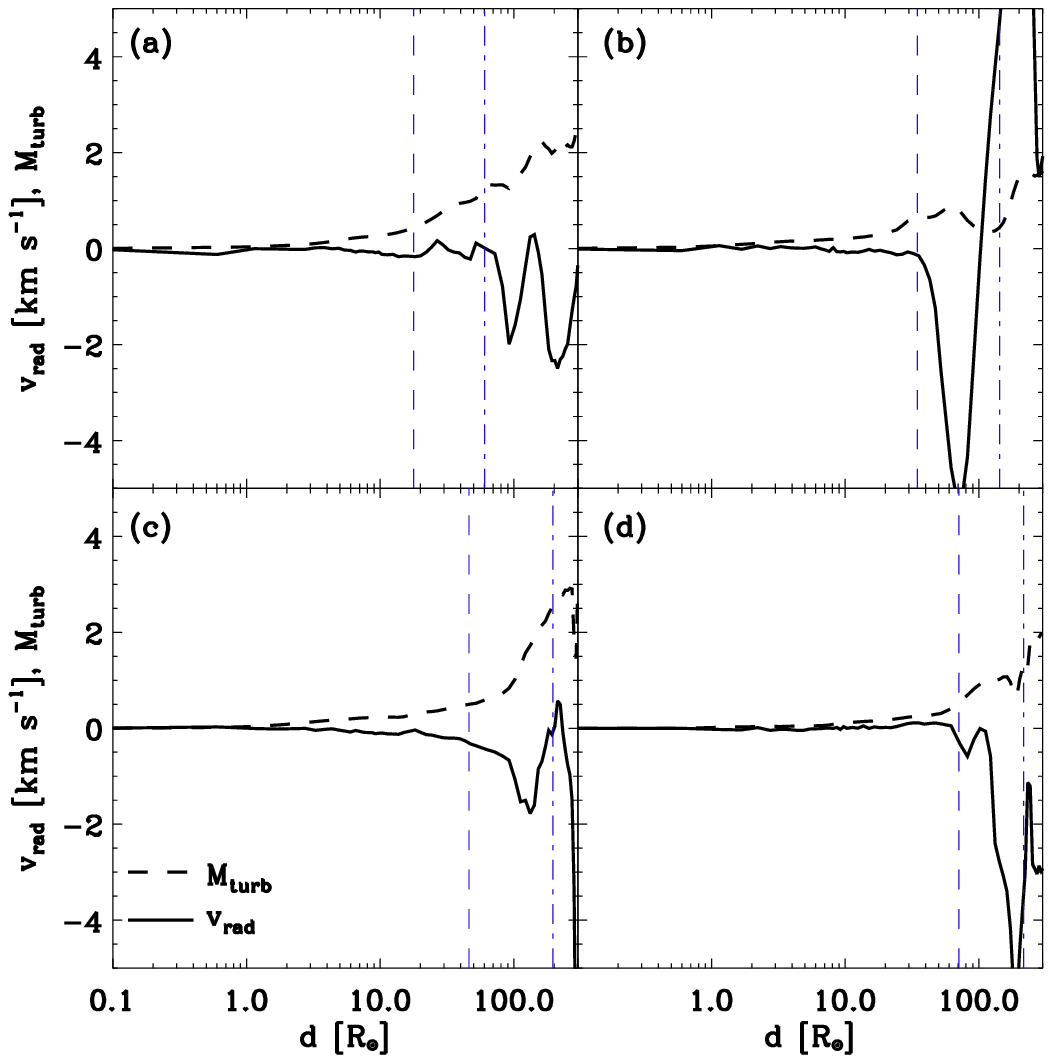}
 \caption{Profile of radial velocity within the most massive protostar of each minihalo (solid lines), measured from the protostellar centers. 
 {\it (a):}  Minihalo 1.
{\it (b):}  Minihalo 2.
{\it (c):}  Minihalo 3.
{\it (d):} Minihalo 4.
Vertical lines employ the same convention as in Figure \ref{vrot}.
 Note that $v_{\rm rad}$ is nearly zero within $R_*$, indicating the hydrostatic equilibrium within the newly formed protostars.
Dashed lines show $M_{\rm turb}$, which reaches supersonic levels outside of $R_{\rm p}$, but is negligible within the protostars.
  }
\label{vrad}
\end{figure*}

A full understanding of the nature of Pop III stars and their impact on the IGM thus requires knowledge of their typical range of spin, and recent observations lend evidence that at least some Pop III stars had significant angular momentum.
For instance, it has been found that some very metal-poor stars in the Milky Way (MW) halo have an anomalous depletion of Li well below the Spite plateau (e.g. \citealt{frebeletal2005,caffauetal2011,bonifacio2012}).  This is sometimes referred to as the `meltdown' of the Spite plateau (\citealt{aokietal2009,sbordoneetal2010}).  Such Li destruction may have occurred through rotationally induced mixing (e.g. \citealt{pinsonneaultetal2002}; see also discussion and references
in \citealt{asplundetal2006}).

Further observations by \cite{chiappinietal2011} also lend evidence for rapid rotation of previous generations of massive stars.  They find anomalous enhancement of Ba, La, and Y within the globular cluster NGC 6522.  These elements may have been produced through an enhanced s-process in rapidly rotating massive stars.  Observations of N/O and C/O abundance ratios in metal-poor stars in the halo as well as damped Lyman-$\alpha$ systems, presented in works such as those of \cite{spiteetal2005} and \cite{pettinietal2008}, can also be more easily explained by enrichment from rapidly rotating massive stars (e.g. \citealt{chiappini2006,hirschi2007}, see also discussion in \citealt{maeder&meynet2012}).

Current observations of O- and B-type stars in our Galaxy reveal that massive stars can indeed be rapid rotators, with rotation rates as high as several tens of percent of break-up speed, up to over 300 km s$^{-1}$ 
(e.g. \citealt{huang&gies2008,wolffetal2008}). Lower-metallicity massive stars such as those within the Magellanic Clouds have also been found to have faster average rotation rates than stars of higher metallicity (e.g. \citealt{hunteretal2008}).  The environments in which these stars formed, however, differs from that of Pop III stars (see, e.g., \citealt{zinnecker&yorke2007}).  While Pop III stars form in DM-dominated minihaloes, massive stars today form within non-DM-dominated giant molecular clouds.  
If a fraction of small Pop III stars were ejected from their minihaloes, however, their accretion would be disrupted and they would remain low-mass and long-lived (e.g. \citealt{greifetal2011}).  If these Pop III stars were later reincorporated into the Milky Way or one of its satellites, we may eventually directly observe them and their rotation rates.  Identifying these as Pop III furthermore requires that they do not accrete too large an amount of metals from the enriched ISM (see, e.g., \citealt{frebeletal2009,johnson&khochfar2011}).  However, such Pop III stars have yet to be found.
To examine the range of possible rotation rates for Pop III stars, and thus their potential for various spin-dependent evolutionary pathways, we therefore utilize numerical simulations, initialized in their proper cosmological context.

 In the work of \cite{stacyetal2011}, we estimated the rotation rate of Pop III stars from a cosmological simulation that resolved scales as small as 50 AU, finding these stars can attain very rapid rotation, potentially as high as their break-up speed.  This was deduced from the rotational and thermal structure of the star-forming disk on the smallest resolvable scales, though the behavior of the gas as it reached the stellar surface and accreted onto the star could not be followed directly.  
 The more recent study by \cite{greifetal2012} employed simulations which reached significantly increased resolution, down to scales of 0.05 R$_{\odot}$, in four different minihaloes.  
They found high rates of mergers between protostars, such that about half of the secondary protostars within a minihalo merge with the primary protostar.  
This was the first simulation to follow fragmentation and merging with this level of resolution, while previous pioneering simulations of comparable resolution were either one-dimensional (e.g. \citealt{omukai&nishi1998,ripamontietal2002}), or three-dimensional but 
unable to follow the gas evolution beyond the formation of the initial hydrostatic core.
We analyze the \cite{greifetal2012} simulations to obtain improved estimates of Pop III rotation rates and internal structure.  Because these calculations resolved scales within the surface of the protostar beginning from several different cosmological realizations, we can now study the effect of mergers on the rotational structure, and the variation of spin for protostars within different host minihaloes.  In Section 2 we give an overview of the numerical methodology used in the simulations.  
In Section 3 we present our results, assessing the main
caveats in Section 4. 
We conclude in Section 5.

 \begin{figure*}
\includegraphics[width=.8\textwidth]{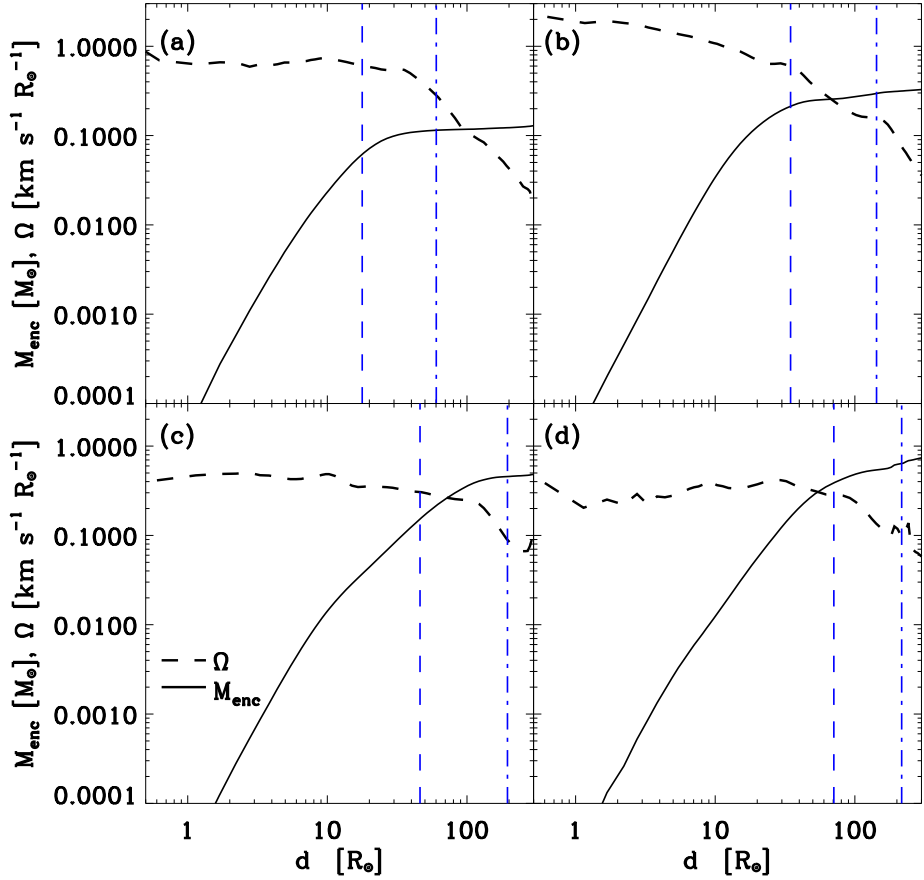}
 \caption{Profile of the enclosed mass $M_{\rm enc}$ in the protostellar regions of each minihalo, as measured from the centers of the protostars (solid lines).  Also shown is the profile of $\Omega$ in units of km s$^{-1}$ R$_{\odot}^{-1}$ (dashed lines). 
Vertical lines have the same meaning as in Figure \ref{vrot}. 
Note the flattening of $M_{\rm enc}$ beyond the edge of each protostar, and the roughly solid-body rotational profiles inside the protostars.
{\it (a):}  Minihalo 1.
{\it (b):}  Minihalo 2.
{\it (c):}  Minihalo 3.
{\it (d):} Minihalo 4.
 }
\label{menc}
\end{figure*}

\section{Numerical Methodology}

\subsection{Initial Setup}

Our work utilizes the simulation output of \cite{greifetal2012}, described in detail therein.  Briefly, the initial conditions for these simulations were taken from \cite{greifetal2011}, in which the calculations were performed with the moving-mesh  code {\sc arepo} (\citealt{springel2010}).  The original cosmological DM simulations employed boxes of lengths 250 and 500 kpc (comoving).  They were initialized with $128^3$ and $256^3$ particles at $z=99$, assuming a $\Lambda$ cold dark matter ($\Lambda$CDM) cosmology with $\Omega_{\rm m} = 0.27$, $\Omega_{\rm b} = 0.046$, $h = 0.71$, and $\sigma_8$ ranging from 0.81 to 1.3.  See table 1 of \cite{greifetal2012} for details.  The simulations are then followed until the formation of the first minihalo with virial mass greater than $5 \times 10^5$ M$_{\odot}$.

\subsection{Cut-Out and Refinement}

Once the site of the first minihalo has been determined, new initial conditions at $z=99$ are generated.  Particles that will become part of the first minihalo, as determined by the original simulations, are replaced by 64 DM particles and 64 mesh-generating points to be used in the hydrodynamic calculations.  Cells and DM particles with increasingly large distances from the high-resolution region are replaced with cells and particles of progressively higher mass. Therefore, regions farther from the target minihalo are more coarsely resolved, reducing the total number of cells and particles to be followed for these refined simulations.

After a cell in the refined simulations reaches a density of 10$^9$ cm$^{-3}$, the central 1 pc is extracted and used as the initial conditions for a further-refined simulation with reflective boundary conditions.  These new simulations are then followed to a density of 10$^{19}$ cm$^{-3}$, after which the central 2000 AU is again extracted for further refinement, such that the final simulations resolve scales as small 0.05~R$_{\odot}$.

\subsection{Chemistry, Heating, and Cooling}

The chemical and thermal network, described in detail in \cite{greifetal2011,greifetal2012}, follows the evolution of H, H$^+$, H$^-$,  H$_2^+$, H$_2$, He, He$^+$, He$^{++}$, D, D$^+$, HD, and free electrons 
(\citealt{glover&jappsen2007}, \nocite{clarketal2011a} Clark et al 2011a).  Particularly important chemothermal processes include cooling through H$_2$ rovibrational transitions at low densities, three-body H$_2$ formation at densities above 10$^8$ cm$^{-3}$, and collision-induced emission (CIE) at densities greater than 10$^{14}$ cm$^{-3}$.  
H$_2$ collisional dissociation cooling also provides an off-set to compressional heating at high densities (e.g., \citealt{omukai2000,yoshidaetal2008}).

\begin{figure*}
\includegraphics[width=.8\textwidth]{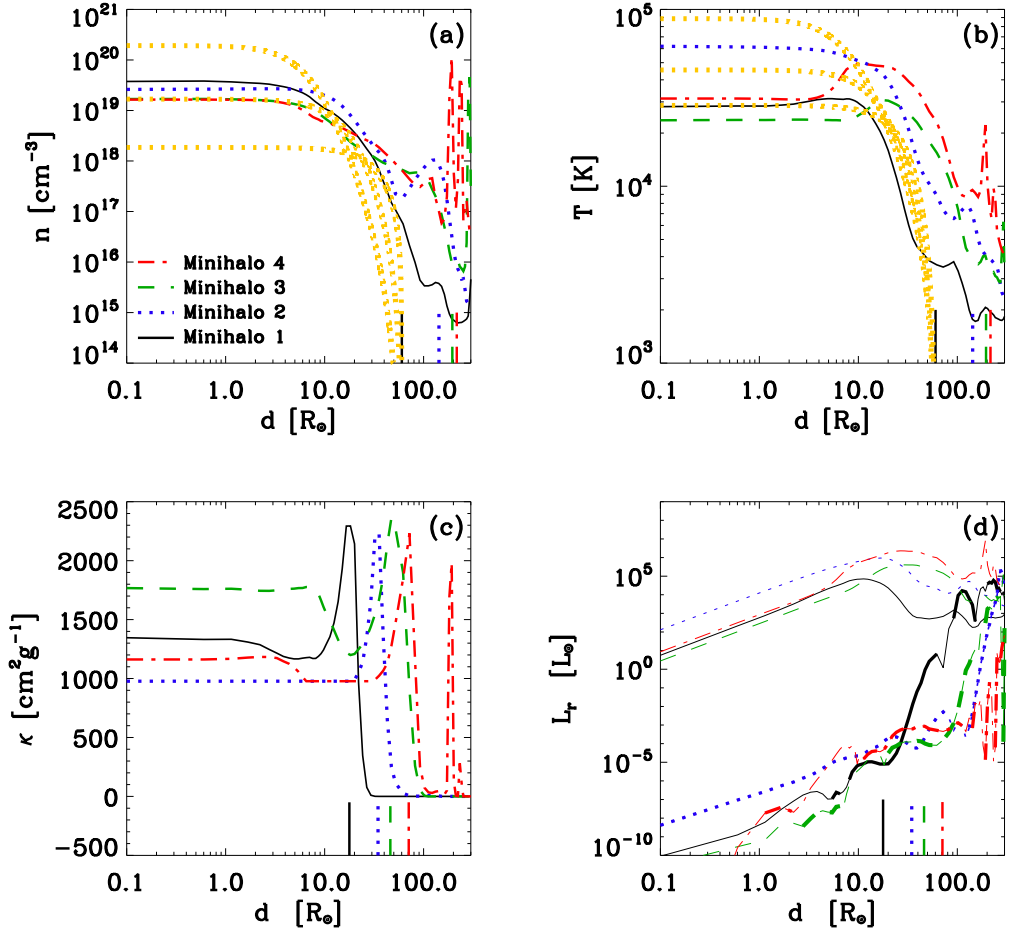}
 \caption{Radial profile of various properties of the most massive protostar in each halo.
{\it (a):}  Number density, 
{\it (b):}  temperature, 
{\it (c):}  opacity $\kappa$ and
{\it (d):} radiative luminosity $L_{\rm r}$
with respect to distance from the protostellar center.  
Black solid line represents the protostar from Minihalo 1, blue dotted line from Minihalo 2, green dashed line from Minihalo 3, and red dash-dotted line from Minihalo 4.
Short vertical lines of corresponding style in panels {\it (a)} and  {\it (b)} show the location of the photospheric surface $R_{\rm p}$.  In panels {\it (c)} and  {\it (d)}, the vertical lines instead denote $R_*$.
Thick yellow dotted lines represent example analytic stellar structure solutions.  From bottom to top, these solutions are for polytropic indices of $n=$1.5, 3.0, and 4.0, with normalizations based on the mass and radius of the Minihalo~1 protostar.  Comparing the solid black line with the yellow dotted lines, the protostellar density is best described by the $n=3$ solution, while the temperature is better-described by the $n=1.5$ solution.  In panel {\it (d)}, upper thin lines correspond to $L_{\rm eff}$.  As expected,  $L_{\rm r}$ approaches $L_{\rm eff}$ near the protostellar surface $R_{\rm p}$.  Thin-lined sections of the $L_{\rm r}$ profiles represent regions where $dT^4/dr$ is positive, leading to negative $L_{\rm r}$ values.  In these sections we thus show the modulus of  $L_{\rm r}$.
Note that at these times the conditions for nuclear ignition have not yet been satisfied.
}
\label{star_vs_r}
\end{figure*}

\subsection{Extraction and Identification of Protostars}

Each protostar, along with its mass $M_*$ and photospheric radius $R_{\rm p}$, was identified using the procedure described in \cite{greifetal2012}.  
In short, each new protostar was found by searching for cells with density surpassing $n=10^{19}$ cm$^{-3}$.  The densest of these cells was used as the center of the candidate protostar.  If the protostellar center was located outside the radius of any previously identified protostars, it was counted as a new protostar.   
The protostellar boundary was defined as the edge of the photosphere, which was calculated by determining the spherically averaged radius where the optical depth reached unity.

Once a candidate cell was located, the optical depth $\Delta\tau$ was determined within $\sim 2\times10^6$ bins.  Bins were found by dividing the area into $N_{\rm ang}\sim 10^4$ uniformly-spaced angular regions that were further divided into $N_{\rm rad}=200$ logarithmically-spaced radial segments between 0.01 and 10 AU, centered on the candidate cell:
\begin{equation}
\Delta\tau_{j, k}=\rho_{j, k}\kappa_{j, k}\Delta r_k,
\end{equation}
where $j$ denotes the angular region, $k$ the radial segment, $\rho_{j, k}$ the mass-weighted density of the bin, $\kappa_{j, k}$ the Rosseland mean opacity, and $\Delta r_k$ the radial extent of the bin.  Opacities were taken from \cite{mayer&duschl2005}.  The integrated optical depth for each bin was then summed along each radius:

\begin{equation}
\tau_{j, k}=\sum_{l=N_{\rm rad}-1}^{l=k}\Delta\tau_{j, l}  \mbox{.}
\end{equation}
\cite{greifetal2012} next calculated the radial index $k_{\rm crit}$ at which the spherically averaged escape fraction,
\begin{equation}
\beta_{{\rm esc}, k}=\frac{1}{N_{\rm ang}}\sum_j\frac{1-\exp{\left(-\tau_{j, k}\right)}}{\tau_{j, k}} \mbox{,}
\end{equation}
drops to $\beta_{\rm crit}=1-\exp{\left(-1\right)}\simeq0.63$, which corresponds to an optical depth of unity. The photospheric radius $R_{\rm p}$ was then set equal to $r_k(k_{\rm crit})$, while $M_*$ was given by the mass enclosed within $R_{\rm p}$.

In our current study, we furthermore define a `hydrostatic' radius $R_*$, defined as the distance from the protostellar center to the radial bin inside $R_{\rm p}$ which has the maximum value of $\kappa$. As will be seen in Section 3.2, the radius $R_* < R_{\rm p}$ where $\kappa$ reaches a peak corresponds more precisely to the location of the protostellar accretion shock.

We focus our study on four protostars, the most massive one found within each of the four minihalos simulated in \cite{greifetal2012}, thereby choosing the protostars most likely to later become massive stars with the greatest impact on their surroundings.
This allows us to probe the early stages of rapidly accreting protostars within a range of host minihalo environments.  

\section{Results}


\begin{figure*}
\includegraphics[width=.8\textwidth]{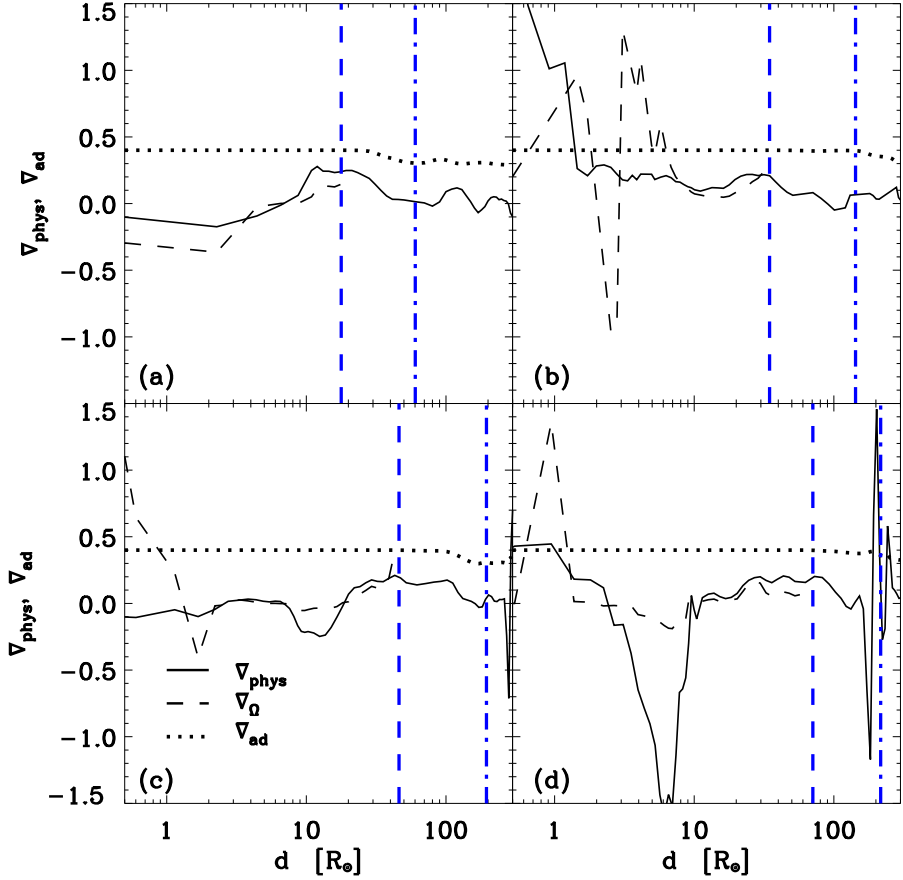}
 \caption{Physical and adiabatic temperature gradients with respect to distance from the center of the protostar, shown at the final snapshot time $t_{\rm fin}$ of each simulation.  Solid lines denote $\nabla_{\rm phys}$, and dotted lines denote $\nabla_{\rm ad}$.  
Dashed lines show $\nabla_{\Omega}$ out to $R_*$, while beyond this radius our approximation for $\nabla_{\Omega}$ no longer applies since the gas is not undergoing solid-body rotation in these regions.
 Vertical dash-dotted lines show the location of the photospheric edge of the protostar at the end of the simulations, while vertical dashed lines denote the location of $R_*$.  
 Convection occurs where  $\nabla_{\rm phys} > \nabla_{\rm ad}  + \nabla_{\Omega}{\rm sin}\theta$.
Thus, these protostars are generally non-convective at these
early times, where the conditions for nuclear ignition have not yet been satisfied.
 }
\label{conv}
\end{figure*}

\subsection{Protostellar Rotation}

\subsubsection{Rotational Profile}

The significant rotational support within the protostars at the end of each simulation can be seen in Figure \ref{vrot}, where we show the radial profile of $v_{\rm rot}$.
For each radial bin we determine its rotational velocity $v_{{\rm rot}}$ as follows:
\begin{equation}
v_{{\rm rot}} $=$ 
\frac {\left[\left(\sum m_i v_{{\rm rot},x,i}\right)^2 +  \left(\sum m_i v_{{\rm rot},y,i}\right)^2 +  \left(\sum m_i v_{{\rm rot},z,i}\right)^2\right]^{1/2} }  {M} \mbox{,}
\end{equation}
where $m_i$ is the mass of a single hydrodynamic cell in the radial bin, $v_{{\rm rot},x,i}$ is the $\vec{x}$ component of cell's rotational velocity, and $M$ is the total mass within the bin.

Figure \ref{vrot} furthermore displays
the Keplerian velocity $v_{\rm Kep}  = \left(G M_{\rm enc} /r\right)^{1/2}$, where $M_{\rm enc}$ is the mass enclosed within distance $r$ from the center of the protostar.
Also shown is an example solid-body rotational velocity profile,
$v_{\rm solid} = \Omega_{\rm max} r$, where $\Omega_{\rm max} = v_{\rm rot,max}/R_{\rm p}$, and $v_{\rm rot,max}$ is the maximum value of $v_{\rm rot}$ found within $R_{\rm p}$.  
At the protostellar surface, the rotational support varies from  
 $\sim$ 80\% of  $v_{\rm Kep}$ for Minihaloes~1 and 3 to $>$ 95\% for Minihaloes~2 and 4.

 In panels {\it b}, {\it c}, and {\it d}, tightly bound secondary protostars are evident as extra peaks in the  $v_{\rm rot}$ profiles beyond the edges of the main protostars.

For comparison, the profile of radial velocity $v_{\rm rad}$ can be seen in Figure \ref{vrad}.  
The large $v_{\rm rad}$ outside $R_*$ signifies the radial inflow onto the protostellar surface, while  $v_{\rm rad}$ falls to zero inside $R_*$ due to the hydrostatic equilibrium of the gas.
We additionally show the turbulent Mach number $M_{\rm turb}$ over a range of radial bins in Figure \ref{vrad} (see also \citealt{greifetal2012}), defined as
\begin{equation}
M_{\rm turb}^2 c_s^2 =\sum_{i} \frac{m_i}{M}\left(v_i - v_{\rm rot }  - v_{\rm rad  }\right)^2 \mbox{,}
\end{equation}
where $c_s$ is the sound speed of the radial bin, $m_i$ is the mass of a cell with index $i$ contributing to the bin, and $M$ is the total mass of the bin.
Turbulence increases where there is radial inflow, and thus is approximately sonic outside of $R_*$ while becoming supersonic outside of $R_{\rm p}$.  However, it is negligible inside of $R_*$ where the gas has attained hydrostatic equilibrium.
 
 \begin{figure*}
\includegraphics[width=.4\textwidth]{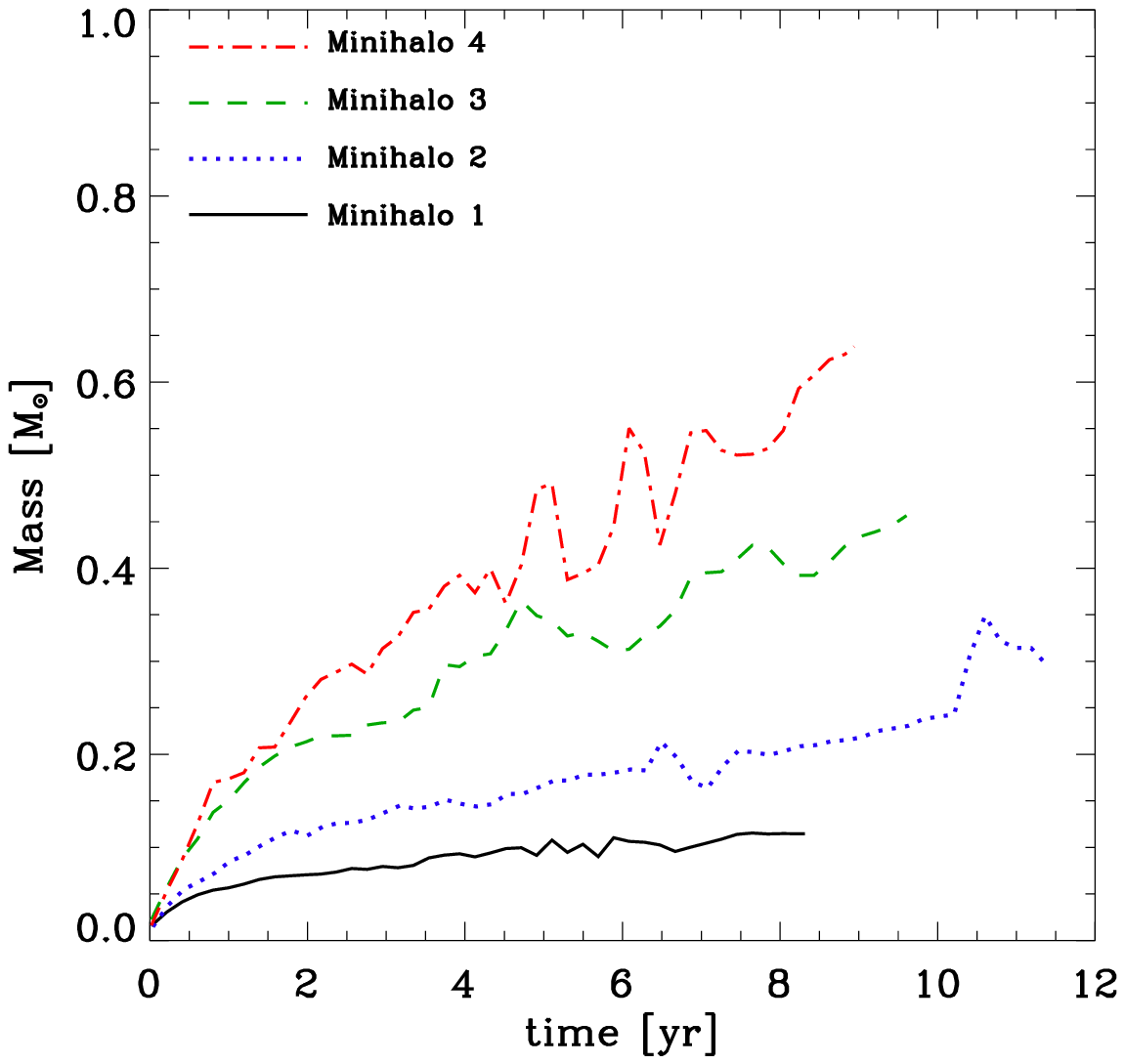}
\includegraphics[width=.4\textwidth]{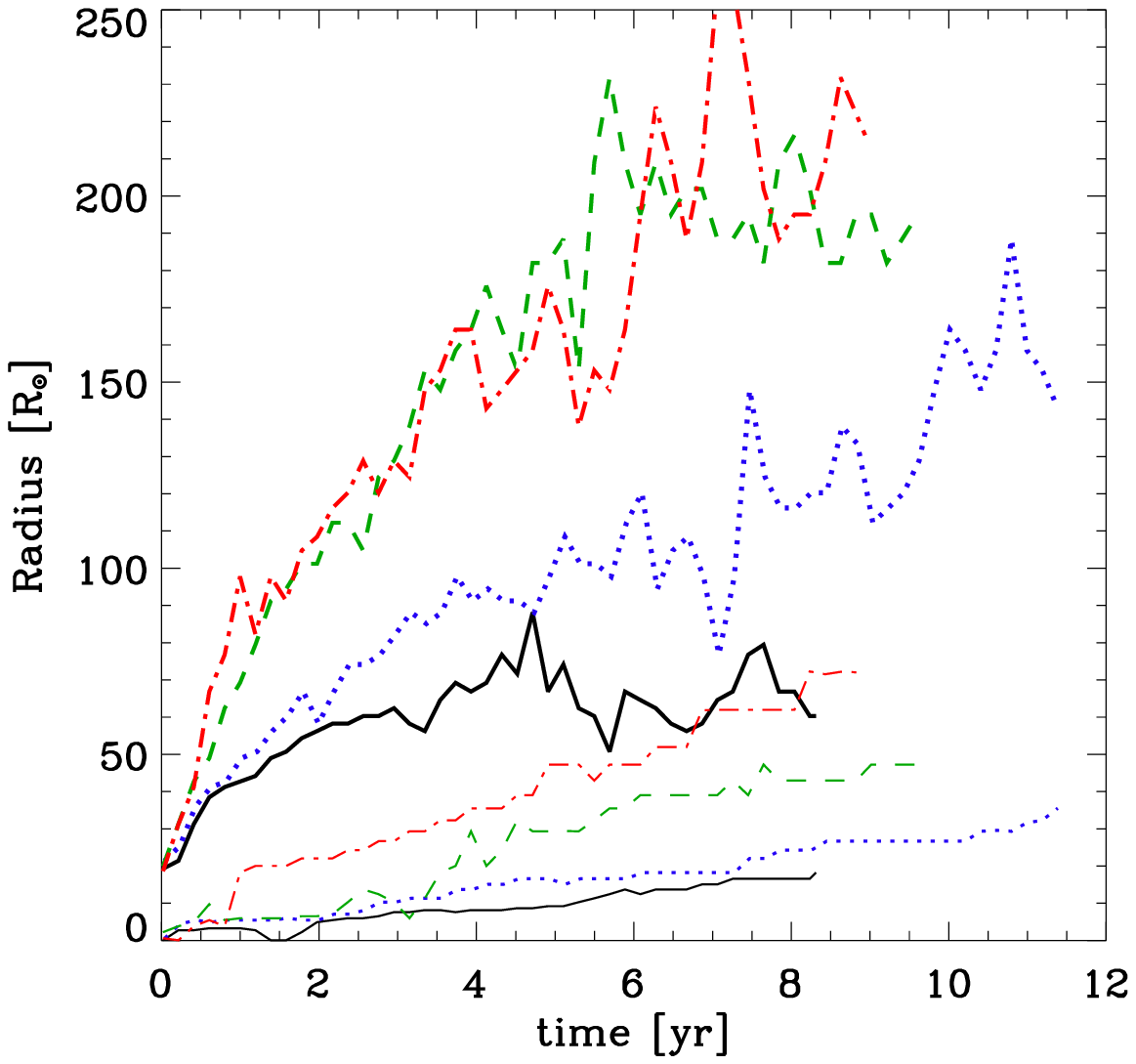}
 \caption{Evolution of various properties of the most massive protostar in each halo.
{\it (Left):}  Mass, 
{\it (Right):}  photospheric radius $R_{\rm p}$ (thick lines) and hydrostatic radius $R_*$ (thin lines), 
over time.  Line styles have the same meaning as in Figure \ref{star_vs_r}.
}
\label{star_vs_t}
\end{figure*}

Within the protostar, $v_{\rm Kep}$ roughly follows $v_{\rm Kep} \propto r$, the same as  $v_{\rm solid}$, due to the nearly constant-density protostellar cores which yield $M_{\rm enc} \propto r^3$.  The decline of $v_{\rm Kep}$ at radii beyond the protostar coincides with the leveling off of the enclosed mass $M_{\rm enc}$ (Fig. \ref{menc}), at which point  $v_{\rm Kep}$ will scale approximately as $r^{-1/2}$ (see also
\nocite{clarketal2011b} Clark et al. 2011b).

Along with $v_{\rm Kep}$, rotation rates within the protostars also follow an approximately solid-body profile.
This is apparent in Figure \ref{menc}, which along with $M_{\rm enc}$ also shows the radial profile of the angular velocity 
$\Omega =  v_{\rm rot}/r$, 
in units of km s$^{-1}$ R$_{\odot}^{-1}$. 
The protostars generally exhibit a near-constant $\Omega$ ranging between $\sim 5 \times 10^{-7}$ to $\ga 1.4 \times  10^{-6}$\,s$^{-1}$. Near the protostellar surface and beyond, $\Omega$ shows the expected drop-off as roughly $\Omega \propto r^{-3/2}$.
For gas within the constant-density cores,
such a profile results from $v_{\rm rot}$ remaining at significant fractions of $v_{\rm Kep}$ throughout the protostar, because gas from the rotationally-supported disk maintains much of its rotational support as it accretes onto the protostar.
Note, however, that the $\Omega$ profiles are not as flat as the inner profiles of models presented in, e.g., \cite{meynet&maeder2000}.  
Small deviation from a solid-body profile within the constant-density core ($d \la 10$ R$_{\odot}$) is due to slight flattening of the protostars even at their inner radii,
as well as angular momentum redistribution provided by torques from turbulent infalling gas and secondary protostars, 
allowing for some differential rotation such that $\Omega = \Omega(r)$. 

\subsubsection{Methods of Angular Momentum Transport}

Aside from these small deviations, perhaps more striking is that the protostars nevertheless maintain roughly solid body rotation even beyond the constant-density cores, nearly out to their photospheric surfaces ($d \sim 100$ R$_{\odot}$), where the density has declined by over an order of magnitude, or several orders of magnitude in the case of the Minihalo 1 protostar.  This implies rapid angular momentum transport on scales of $R_{\rm p}$, which is driven mainly by sonic turbulence.  Figure \ref{vrad} shows that between $\sim$ 10 and 100 R$_{\odot}$ the protostellar gas transitions from non-turbulence to roughly sonic turbulence.  The timescale for this hydrodynamic transport can be estimated as

\begin{equation}
t_{\rm trans} \sim R_{\rm p}/c_s \mbox{,}
\end{equation} 

\noindent where $R_{\rm p} \sim 100$ R$_{\odot}$ and $c_s \sim 5$ km s$^{-1}$.  This yields $t_{\rm trans} \sim 1$ yr. Turbulent transport is therefore sufficiently rapid to account for the majority of angular momentum redistribution within the outer regions of the protostars during the time covered by the simulation.  
On larger scales of the protostellar disk, additional angular momentum transport occurs through gravitational torques generated by, e.g., spiral arm patterns and secondary protostars, as well as dynamical friction.  This allows for protostellar mergers to occur on nearly free-fall timescales (see discussion in \citealt{greifetal2012}).

Angular momentum transport through unphysical numerical diffusion is negligible.  As described in \cite{springel2010}, the Langragian scheme used in {\sc arepo} generates significantly less numerical diffusivity than corresponding Eulerian methods.  For instance, when {\sc arepo} was tested for its accuracy in solving the Gresho vortex problem (section 8.5 of \citealt{springel2010}), in which rotating vortex motion is applied to constant-density gas, minimal error was generated ($\la$ 1\% for intermediate and high resolution cases, converging to $\sim$ 0.1\% for the highest resolution case).  These low errors were maintained even when the vortex was set in motion through the box, unlike corresponding tests with Eulerian schemes that displayed increased error due to numerical diffusivity.  {\sc AREPO} thus demonstrates good conservation of angular momentum and vorticity with minimal numerical diffusion effects.

Within stars that are not undergoing accretion or mass loss, redistribution of angular momentum can also occur through processes such as shear stress, meridional circulation, and convection (e.g., \citealt{meynet&maeder2000}).  
Eddington-Sweet circulations occur in radiative regions of a star on the timescale $t_{\rm ES}$.  In a uniformly rotating star this is approximately

\begin{equation} 
t_{\rm ES} = \frac{GM_*^2}{L_*R_*}\frac{GM_*}{\Omega^2 R_*^3} = t_{\rm KH} \frac{GM_*}{\Omega^2 R_*^3}
\end{equation}

\noindent (e.g., \citealt{maeder&meynet2012}), where $t_{\rm KH}$ is the Kelvin-Helmholtz timescale.   $M_*$, $R_*$, and $L_*$ are the mass, radius, and luminosity of the protostar.  Typical values for $M_*$ and $R_*$ at the end of this simulation are  $\sim$~0.5 M$_{\odot}$ and  
50~R$_{\odot}$, while $R_{\rm p}$ is a few times larger ($\ga$ 100 R$_{\odot}$). 
The mass-weighted value of the radiative luminosity $L_r$ within $R_{\rm p}$, for regions where $L_r$ is positive, ranges from $\sim 10^{-3}$ L$_{\odot}$ to $1$~L$_{\odot}$ (see also following section).  
Average $L_r$ values within $R_*$ are smaller and range from  $\sim$ $10^{-5}$ L$_{\odot}$ to $10^{-3}$ L$_{\odot}$ since the simulations are still in early pre-nuclear ignition stages.
If we use larger photospheric values to estimate $t_{\rm ES}$,  1 L$_{\odot}$ and 200 R$_{\odot}$,
we then find typical $t_{\rm ES}$ values of $\sim$ 1000 yr, well in excess of the simulated time. For the much smaller luminosities encountered in the interior even longer times would result.

Note that for stars that have contracted to the main-sequence (MS), $t_{\rm ES}$ will typically be several orders of magnitude longer than our estimate, 
e.g., $t_{\rm ES} \sim t_{\rm KH} \sim 3 \times 10^7$ yr for a critically rotating Sun-like star, and longer if its rotation rate is slower.
Also note that our estimate for $L_*$ is several orders of magnitude lower than what might be expected in the outer regions of the star (see following section), or from  the accretion luminosity $L_{\rm acc}$:

\begin{equation}
L_{\rm acc} = \frac{G M_* \dot{M}}{R_*} \mbox{.}
\end{equation}

\noindent  For $\dot{M} \sim 5 \times 10^{-2}$ M$_{\odot}$ yr$^{-1}$, our typical protostar will have $L_{\rm acc} \sim 4000$ L$_{\odot}$.  These high luminosity values, however, are not representative of the radiative flux through the majority of the protostellar interior, so we use the lower values of $L_*$ quoted above.


In contrast to Eddington-Sweet circulation, shear stress and convective mixing are dynamical processes that occur on short timescales of a fraction of a year ($\la0.1$ yr, e.g. \citealt{zahn1992}). 
However, as we will describe in Section 3.3, we do not find significant convective instability in the protostars at the early times simulated.  The requirement for shear instability can be expressed by the
Richardson criterion:

\begin{equation}
Ri \equiv \frac{g_{\rm grav}}{\rho} \frac{d\rho/dz}{\left(dV/dz\right)^2} < Ri_{\rm crit} \mbox{,}
\end{equation}

\noindent where $g_{\rm grav}$ is the gravitational acceleration, $dV/dz$ is the change in horizontal velocity between two layers (i.e., spherical shells), and $Ri_{\rm crit}$ is usually taken as equal to 1/4 (e.g. \citealt{maeder&meynet2012}).  At distances of $\sim$ 100 R$_{\odot}$ from the protostellar center, $g_{\rm grav} \sim 1$ cm s$^{-2}$, $n \sim 10^{17}$ cm$^{-3}$, and $V \sim 10$ km s$^{-1}$.  Taking the changes in $\rho$ and $V$ between the distances of 10 and 100 R$_{\odot}$, we find that $Ri\sim 10-100$.  We also confirmed the accuracy of this estimate with a more precise calculation from the simulation snapshots.  Thus, the density gradient strongly suppresses the shear instability. Thermal diffusivity may have an opposing destabilizing effect, but only on thermal timescales longer than followed in our simulations.  

In short, angular momentum is transported through turbulence and gravitational torques, and angular momentum inflow through the disk provides the protostars with substantial rotational support.
These simulations confirm the prediction of \cite{stacyetal2011}, based on approximate modeling of the sub-grid physics, that the significant rotational support found in the large-scale $\sim$1000 AU star-forming disk down to a few tens of AU would also persist on unresolved AU and stellar scales. Furthermore, our analysis demonstrates that the high rate of protostellar merging does not lead to significant transfer of angular momentum away from the main protostars.

\begin{figure*}
\includegraphics[width=.8\textwidth]{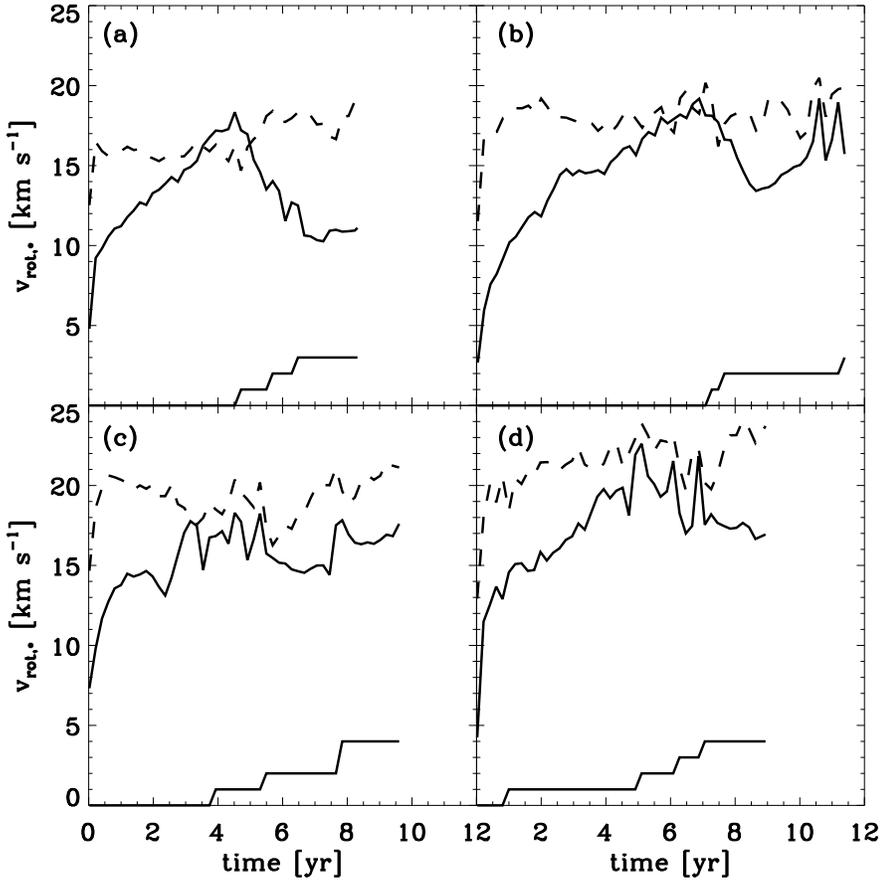}
 \caption{Rotational velocity $v_{\rm rot,*}$ over time for each of the main protostars. Dashed lines show the evolution of $v_{\rm Kep,*}$ for comparison. Bottom lines of each panel also show the number of mergers undergone by the main protostar over time.  The stars maintain a rotational velocity that is a significant fraction of  $v_{\rm Kep,*}$, and the Minihalo 1 star even temporarily has  $v_{\rm rot,*} > v_{\rm Kep,*}$, though this is an artifact of our spherical averaging scheme and the protostar's distortion during its initial merger event.  }
\label{vrot_vs_t}
\end{figure*}

\subsection{Internal Structure}

\subsubsection{Density and Temperature Profiles}

In Figure \ref{star_vs_r}, we show radial profiles of additional quantities for the most massive protostar in each minihalo, taken at the final simulation output.  
We compare the density and temperature structure of the protostar of Minihalo 1 to a range of polytropic models (yellow dotted lines in Fig. \ref{star_vs_r}), where

\begin{equation}
\rho(r) = \rho_{\rm c} \, \theta^n (r) \mbox{,}
\end{equation}

\noindent and

\begin{equation}
T(r) = T_{\rm c} \, \theta (r) \mbox{.}
\end{equation}

\noindent $\theta(r)$ is determined through solving the Lane-Emden equation, which describes the relation between $\theta(r)$ and the dimensionless radius $\epsilon(r) = r/r_n$ depending upon the polytropic index $n$ (see, e.g., \citealt{hansenetal2004}.) 
The normalization factors $r_n$, $\rho_{\rm c}$, and $T_{\rm c} $ are determined by the stellar mass and 
photospheric radius, $M_*$ and $R_{\rm p}$:
\begin{equation}
r_n = R_{\rm p}/ \epsilon_1 \mbox{,}
\end{equation}

\begin{equation}
\rho_{\rm c}   = \left( \frac{\overline{\rho}}{3} \right) \left( \frac{\epsilon_1} {-\theta'_1}  \right)\mbox{,}
\end{equation}

\begin{equation}
\overline{\rho} = \frac{M_*} { \frac{4}{3} \pi R_{\rm p}^3}  \mbox{,}
\end{equation}

\begin{equation}
T_{\rm c}  = \left( \frac{1}{n+1}  \right)  \left( \frac{1}{-\epsilon_1\theta'_1} \right)  \left(\frac{G \mu m_{\rm H}} { k_{\rm B}}  \right) \left( \frac{M_*}{R_{\rm p}} \right)  \mbox{,}
\end{equation}

\noindent where $\epsilon_1$ and $\theta'_1$ are the values of $\epsilon$ and 
$d\theta / d\epsilon$ 
at $\theta(\epsilon_1)=0$.  The mass of a hydrogen atom is $m_{\rm H}$, $k_{\rm B}$ is Boltzmann's constant, and $\mu$ the mean molecular weight. 
We take the mean molecular weight to be $\mu=1.22$, which is the average molecular weight of all cells with $n>5\times10^{19}$ cm$^{-3}$.   We note that $\mu$ is not constant throughout the protostar, contributing to imperfect fits between the simulation and the polytropic solutions.  
However, the protostar of Minihalo 1 seems to best follow models with $n$ ranging from 1.5 to 3.  This applies to the main protostars of the other minihaloes as well, though we do not display these fits in Figure \ref{star_vs_r}.

\subsubsection{Opacity and Luminosity Profiles}

The Rosseland mean opacity $\kappa$, shown in panel {\it (c)} of Fig.~\ref{star_vs_r}, is determined as described in \cite{greifetal2012}. The peak in each $\kappa$ curve coincides with the protostellar accretion shock, roughly corresponding to where the temperature falls below $\sim 2 \times 10^4$ K and density below $\sim 10^{18}$ cm$^{-3}$, with the opacity rapidly dropping at radii beyond the shock.  
The   $\kappa$  peaks mark where the predominant opacity contribution transitions from H bound-free (bf) absorption to 
H$^-$ bf absorption as the gas phase converts from ionized to neutral (e.g. \citealt{mayer&duschl2005}).  At larger radii beyond $\sim$ 100 R$_{\odot}$, the main contribution to opacity is from H$_2$.

\begin{figure*}
\includegraphics[width=.8\textwidth]{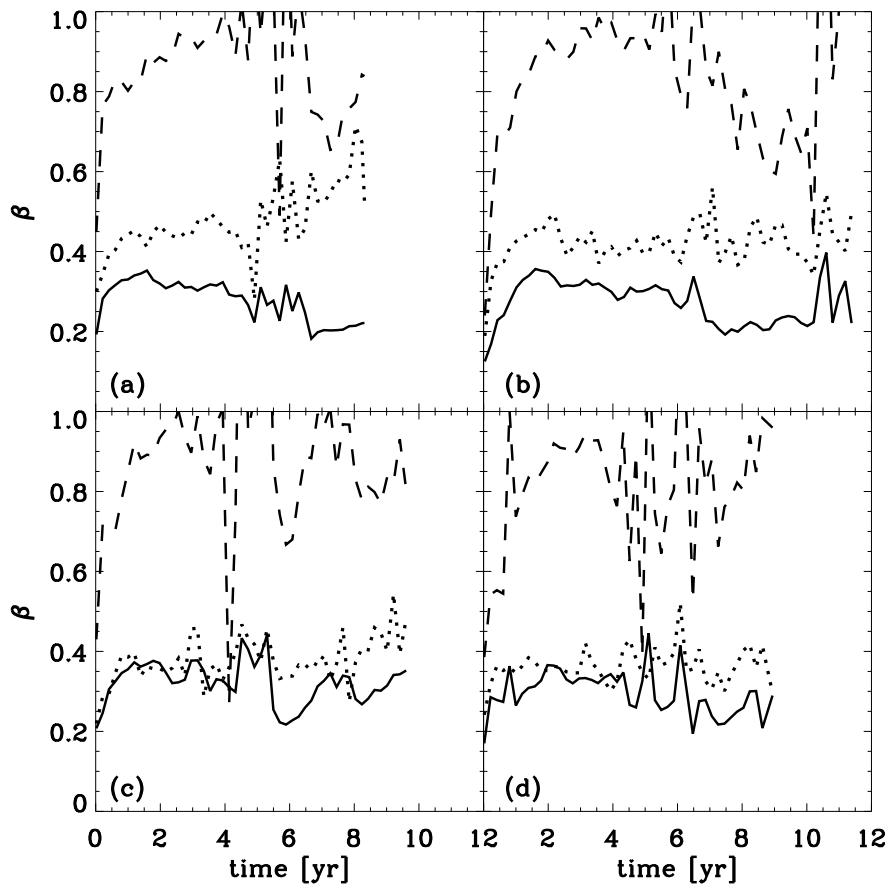}
 \caption{Level of rotational support $\beta$ over time for each of the main protostars (solid lines).  Dotted lines show the level of rotational support of the gas surrounding the protostar out to distances of 10$R_{\rm p}$.  
Dashed lines show the ratio of $v_{\rm rot}$ to $v_{\rm Kep}$ at the protostellar surface $R_{\rm p}$.
After the first $\sim 1-2$\,yr, $\beta$ fluctuates around a nearly constant value of 0.3.}
\label{beta_vs_t}
\end{figure*}

Panel  {\it (d)} of Fig. \ref{star_vs_r} estimates the radiative luminosity profile $L_r$ of each star, defined as

\begin{equation}
L_r\left(r  \right) = -\frac{4\pi a \, c r^2}{3 \kappa \rho} \frac{dT^4}{dr}
\end{equation}

\noindent where $a$ is the radiation constant, $r$ is the distance from the stellar center, and $dT^4/dr$ is taken directly from the simulation output.  The above equation describes the luminosity at a given radius using a diffusion approximation to radiative transfer and assuming that all energy is transported through the protostar by radiation.  In this case the protostellar luminosity $L_*$ is equal to $L_r$.  The simulation, however, did not include full modeling of radiative transfer, and instead assumed a local escape fraction using the Sobolev approximation.  The resulting $L_r$ profile should be interpreted only as an approximation.  For comparison purposes, panel ({\it d}) of Figure \ref{star_vs_r} also shows $L_{\rm eff}(r) = 4 \pi r^2 \sigma_{\rm SB} T^4(r)$. As expected, for each protostar $L_r$ approaches $L_{\rm eff}(r) $ near $r=R_{\rm p}$, where $T$ approaches the effective temperature $T_{\rm eff}$.  We also point out that regions of the star where $dT^4/dr$ is positive lead to negative values of $L_r$.  Regions with negative  $L_r$ values correspond to the thin-lined sections of the profiles, where instead of the negative $L_r$ values we have plotted the modulus.

The accretion flow outside of the photospheric surface $R_{\rm p}$ is subject to supersonic turbulence (see Fig. \ref{vrad}), which provides an additional source of heating.  The values of $L_{\rm r}$ just beyond $R_{\rm p}$ are typically on the order of 1000 L$_{\odot}$, similar to the value $L_{\rm acc} \sim 4000$ L$_{\odot}$ quoted in the previous section. The main contribution to the luminosity is thus the impact of accreting gas onto the protostellar surface, though there is some contribution from the turbulence generated by the inflow as well.  For $M_{\rm turb} = v_{\rm turb}/c_s \sim 2$, and $c_s \sim$ 5 km s$^{-1}$, the turbulent velocity $v_{\rm turb}$ is 10 km s$^{-1}$.  Referring to Figure \ref{menc}, the turbulent region beyond $R_{\rm p}$ (e.g., the region between 200 and 300 R$_{\odot}$ for the protostar of Minihalo 2) contains a small mass of approximately 0.01 M$_{\odot}$, yielding a kinetic energy of 10$^{43}$ erg.  With typical radial velocities of 3 km s$^{-1}$, the timescale for gas to cross the turbulent region is $\sim 10^7$ s, yielding an energy production rate of 10$^{36}$ erg s$^{-1}$, or $\sim$ 250 L$_{\odot}$.  While not negligible, this is significantly smaller than $L_{\rm acc}$ 

We furthermore point out that the effects of radiation pressure were not accounted for in the simulations.  We expect radiation pressure from direct ionization to be a negligible effect, as the radially distended protostars have a typical effective temperature of 
$\sim$ 3000-4000 K, assuming they have surface luminosities of $L_* = L_{\rm acc} \sim$ 4000 L$_{\odot}$.  Note that this agrees well with the gas temperature at the photospheric surface of the protostars, $\sim$ 3000 - 5000 K (see panel {\it b} of Fig.~\ref{star_vs_r}).  These protostars thus do not yet ionize their surrounding gas.

For gas within the protostar, we can estimate the radiation pressure provided due to $L_r$ using

 \begin{equation}
 \frac{dP_{\rm rad}}{dr} =  -\frac{\kappa \rho L_r}{4\pi r^2c}   \mbox{\ ,}
 \end{equation}

 


\noindent However, because the temperatures within the protostars are better-determined than our estimates for $L_r$, we instead approximate $P_{\rm rad}$ by assuming local thermodynamic equilibrium and $P_{\rm rad} = (1/3)a T^4.$  Note that inputting this relation into Equation 17 and solving for $L_r$ yields our Equation 16.
We estimate the radiation pressure just within the photospheric surface, usually at $r \sim 100$~R$_{\odot}$, where typical temperatures are $\sim$ 4000 K, and we find $P_{\rm rad} \sim 1$  dyne~cm$^{-2}$.
The radiation pressure exerted upon the outer atmosphere of the protostars is thus negligible compared to the thermal pressure, $P_{\rm therm} = nk_{\rm B}T \sim 5 \times 10^4$~dyne~cm$^{-2}$ for representative values of $n=10^{17}$\,cm$^{-3}$ and $T=4000$\,K.

Inside the protostar, 
typical gas temperatures and densities in the inner protostellar regions are $T \sim 3 \times 10^4$\,K and $n \sim 3 \times10^{19}$\,cm$^{-3}$, yielding a thermal pressure of $P_{\rm therm} \sim 10^8$ dyne~cm$^{-2}$.  This is again significantly greater than the radiation pressure, 
$P_{\rm rad} \sim 2000$  dyne~cm$^{-2}$.


We can also estimate the strength of radiation pressure due to Thomson scattering by comparing $L_r$ within the ionized region of the protostar to the Eddington luminosity, $L_{\rm Edd} = 4\pi G M_* m_{\rm H} c/\sigma_{\rm T}$, where $\sigma_{\rm T}$ is the Thomson scattering cross section.  
Our typical protostar has $L_{\rm Edd} \sim 10^4$ L$_{\odot}$, orders of magnitude higher than $L_r$ within the ionized gas.  Radiation pressure will thus not be a significant effect until nuclear burning has commenced.

\begin{figure*}
\includegraphics[width=.4\textwidth]{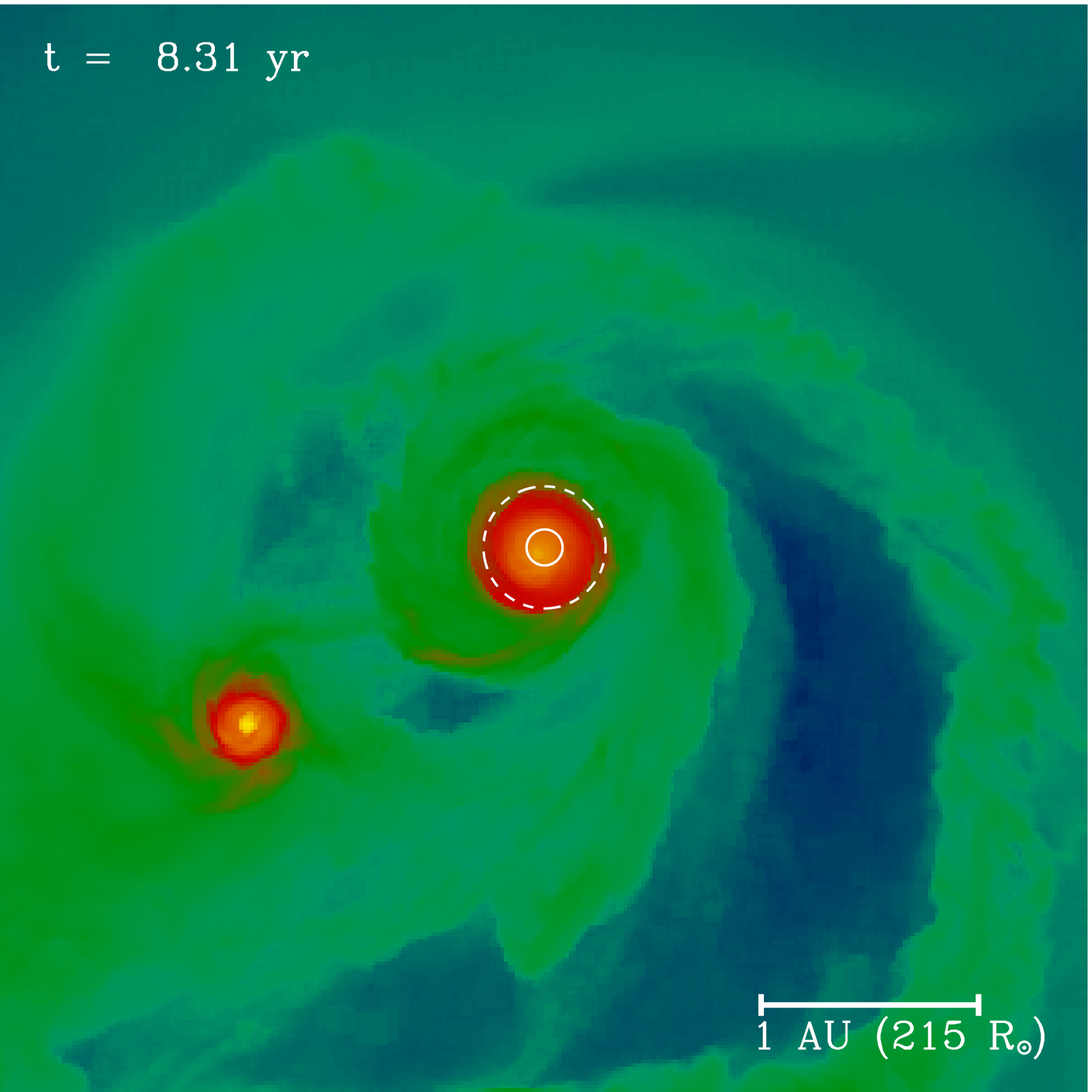}
\includegraphics[width=.4\textwidth]{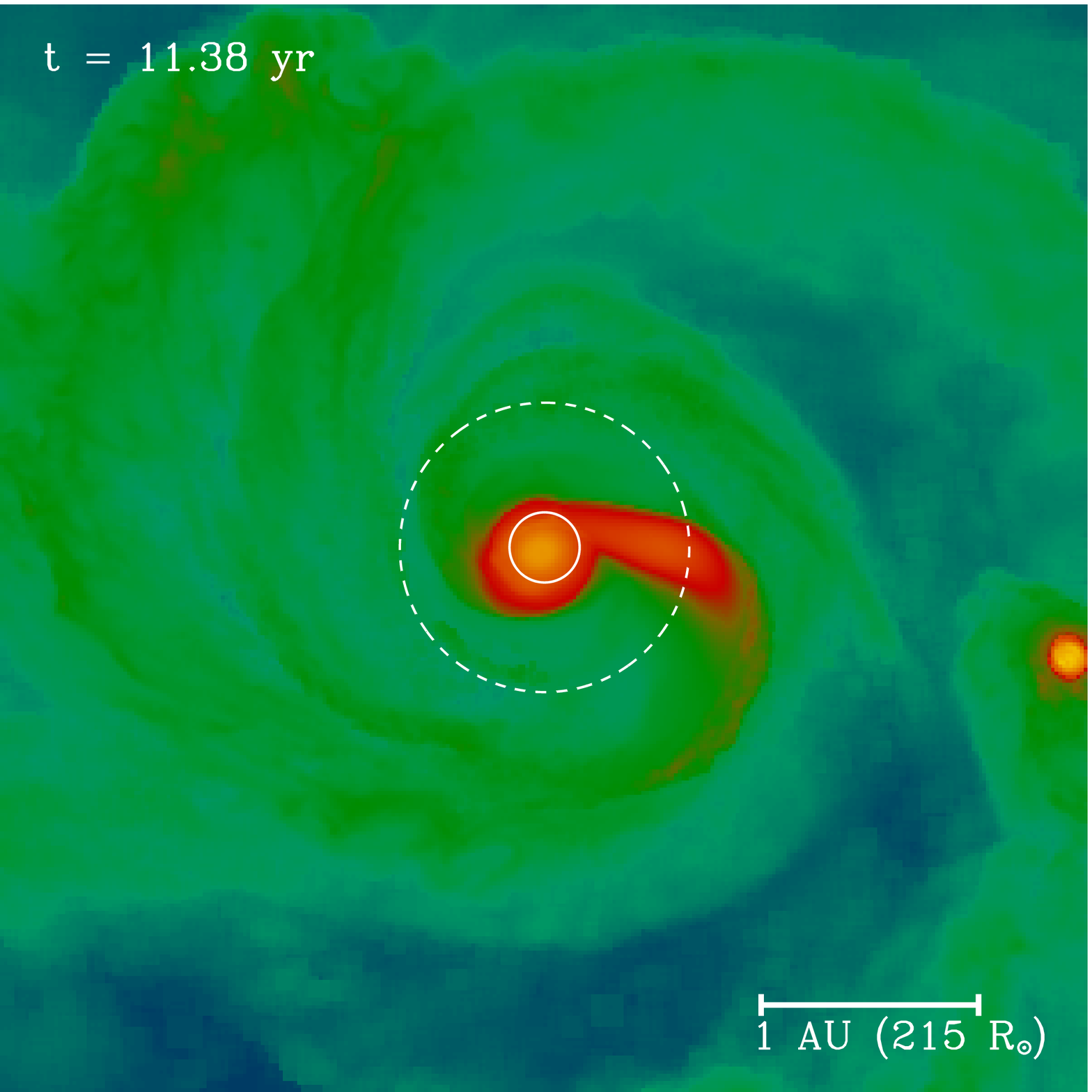}
\includegraphics[width=.4\textwidth]{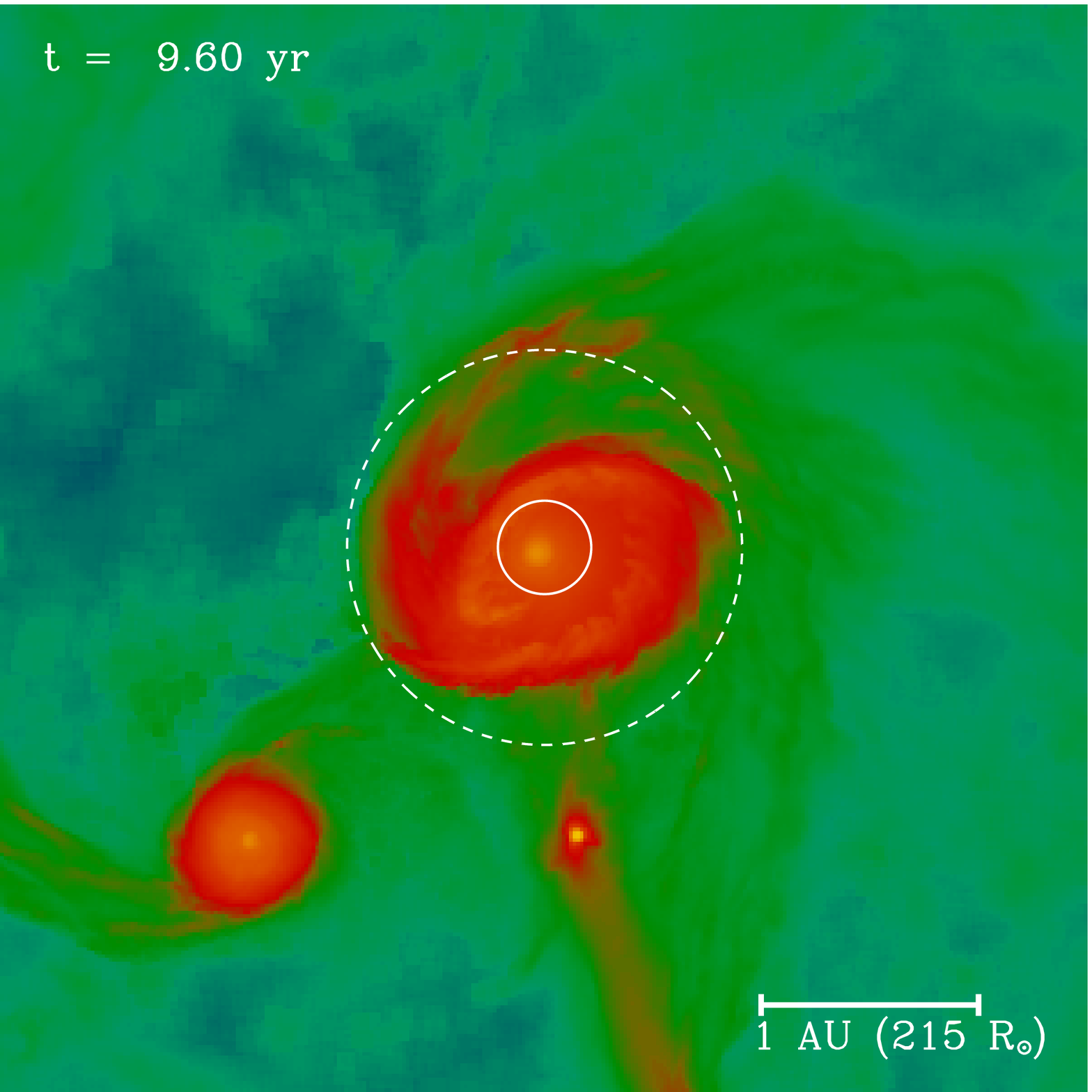}
\includegraphics[width=.4\textwidth]{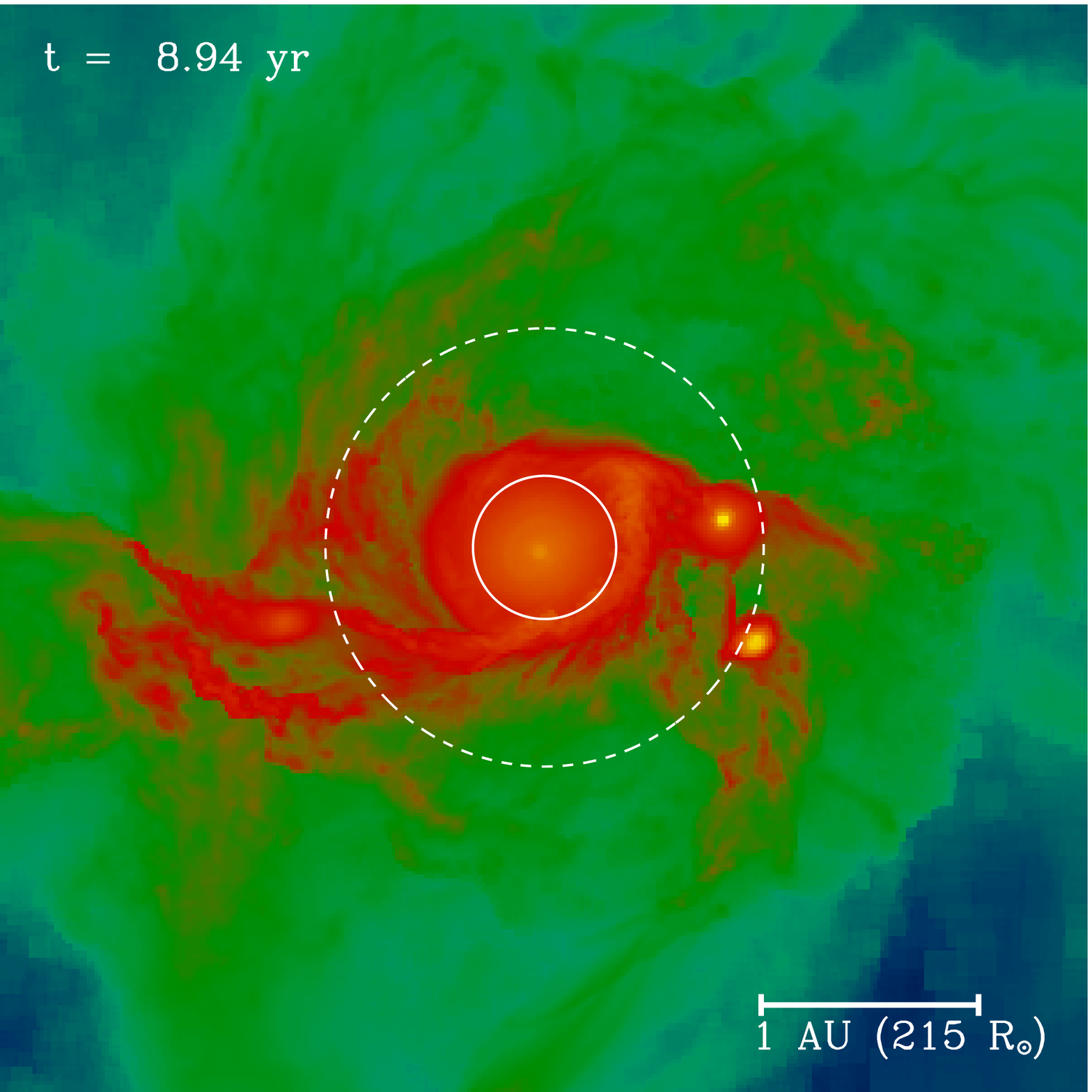}
 \caption{Morphology of the main protostar within each minihalo, 
 where images are oriented to show the face of the disk.    
Each panel has a width of 5 AU.  A line of length 1 AU is also shown in each panel for comparison.  
Dashed white circles depict the size of the photosphere of the main protostars, while solid white circles denote $R_*$.
 The color scale ranges from density of $10^{12}$~cm$^{-3}$ to $10^{21}$~cm$^{-3}$.  
The diskiness and spiral structure of the accreting gas is readily apparent. 
}
\label{star_morph}
\end{figure*}

\begin{figure*}
\includegraphics[width=.4\textwidth]{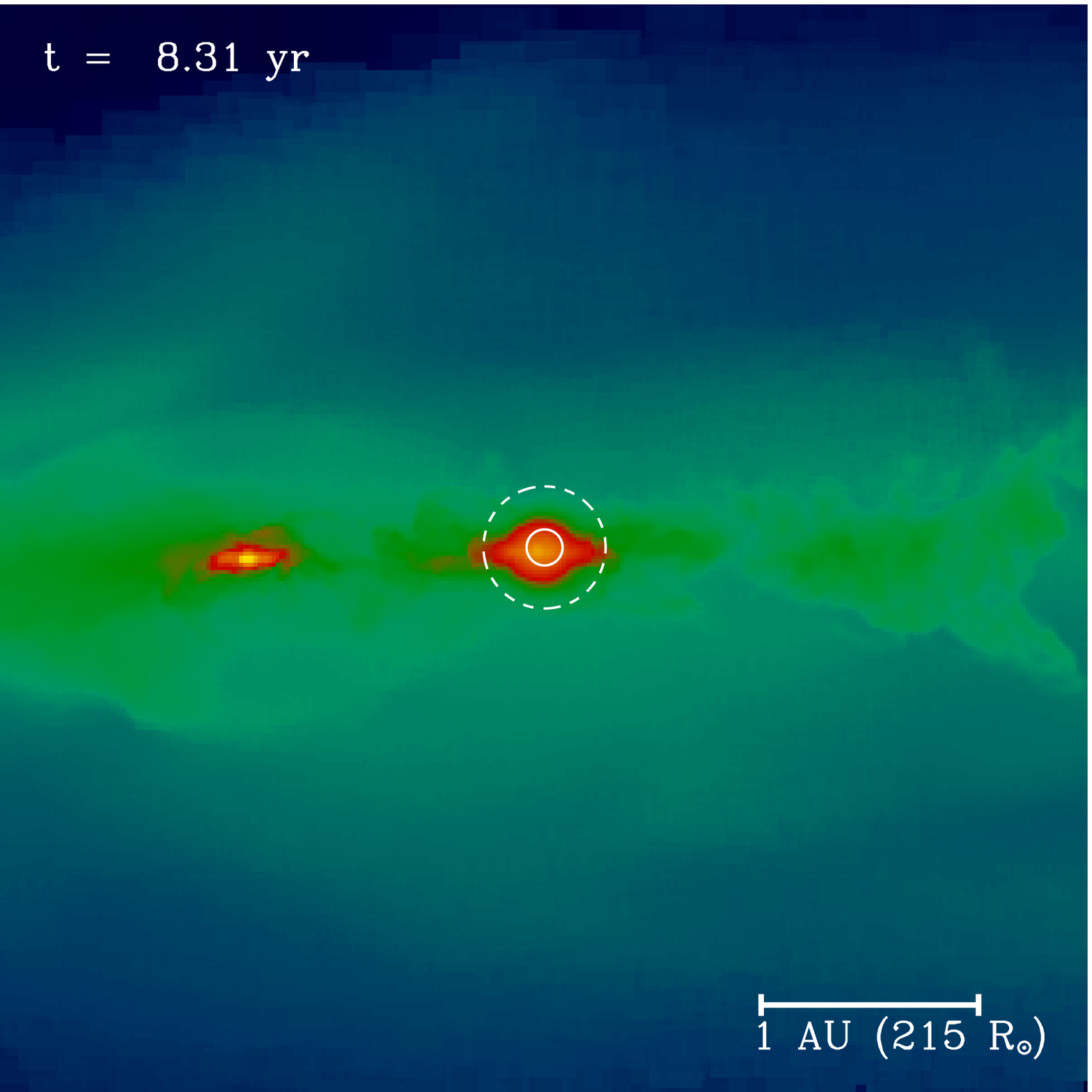}
\includegraphics[width=.4\textwidth]{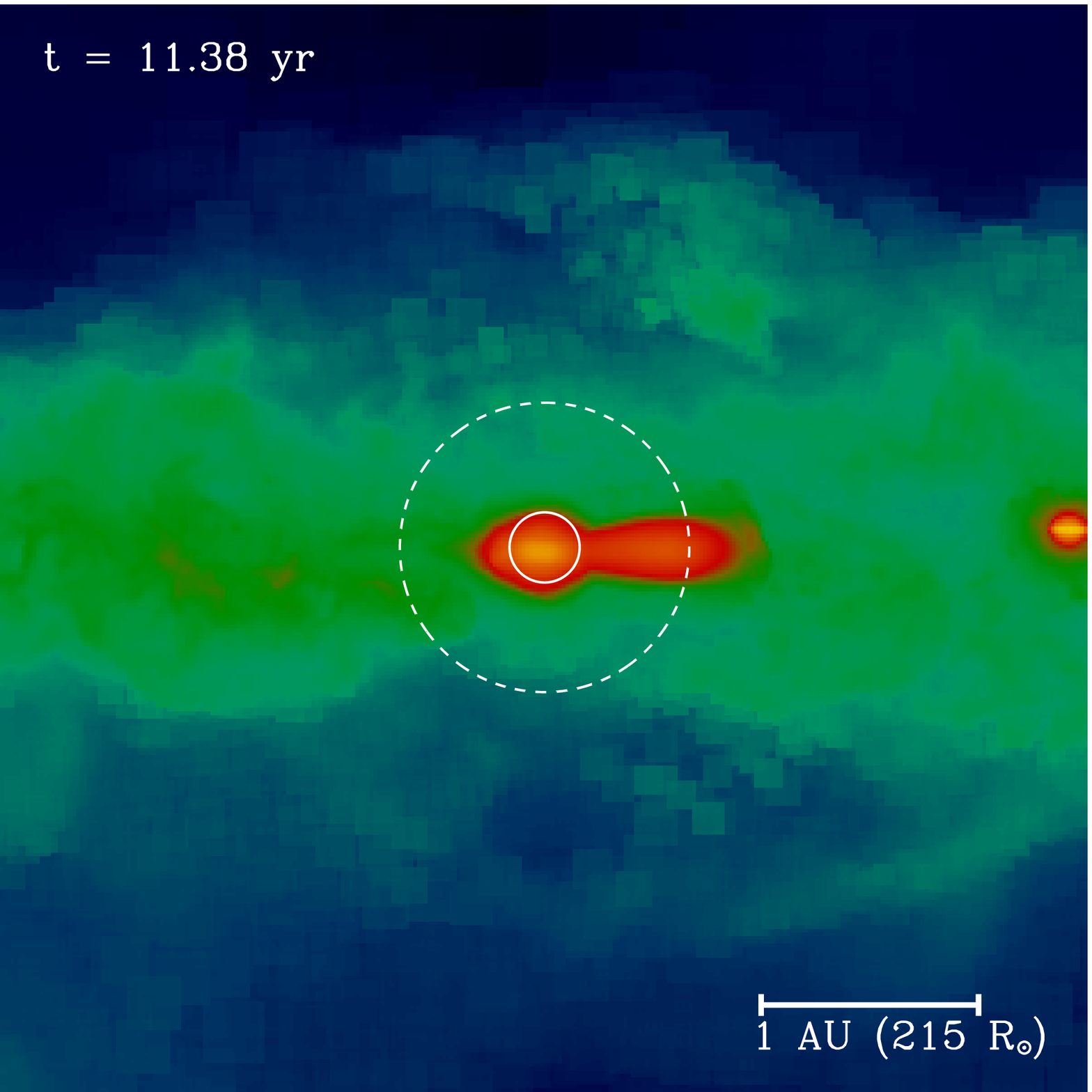}
\includegraphics[width=.4\textwidth]{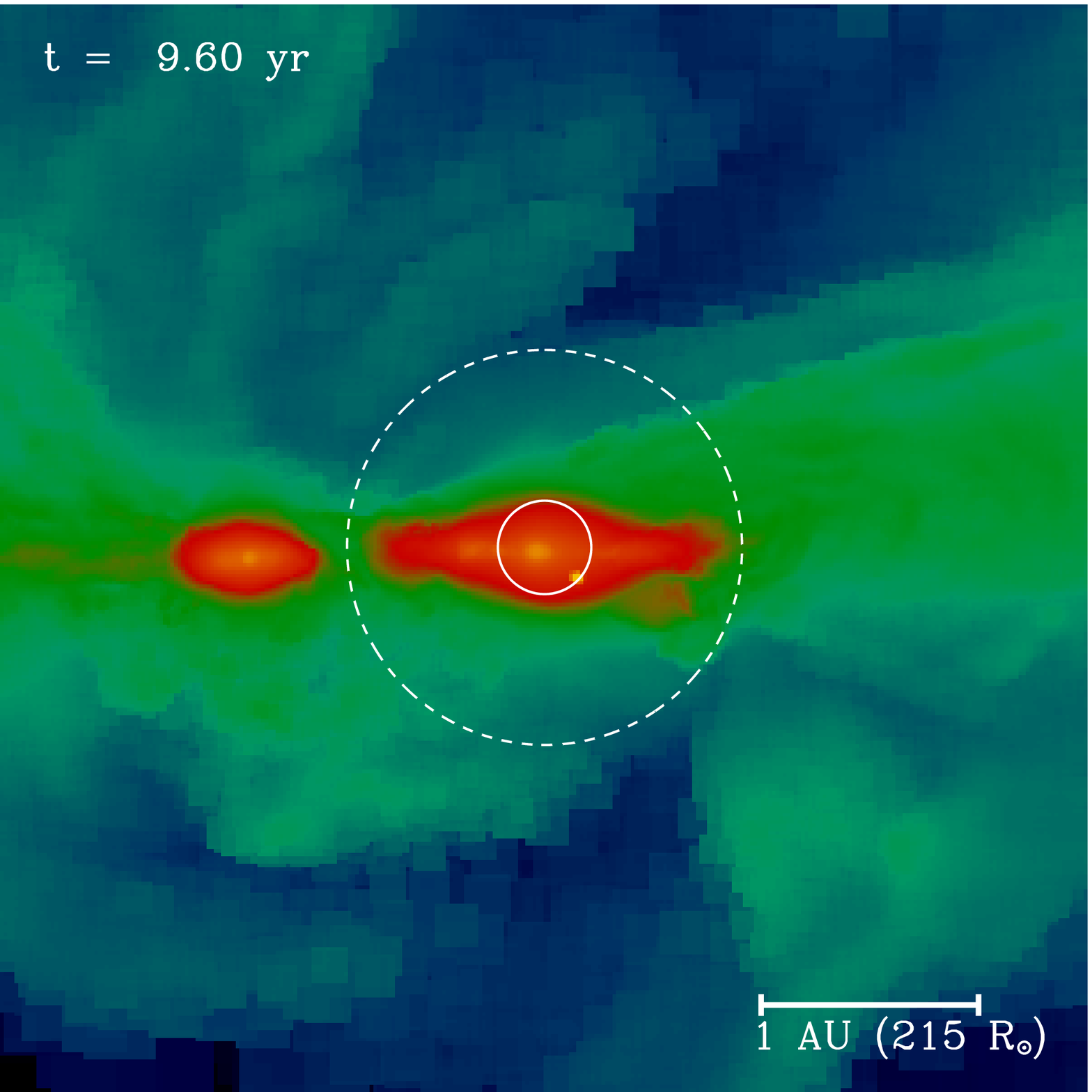}
\includegraphics[width=.4\textwidth]{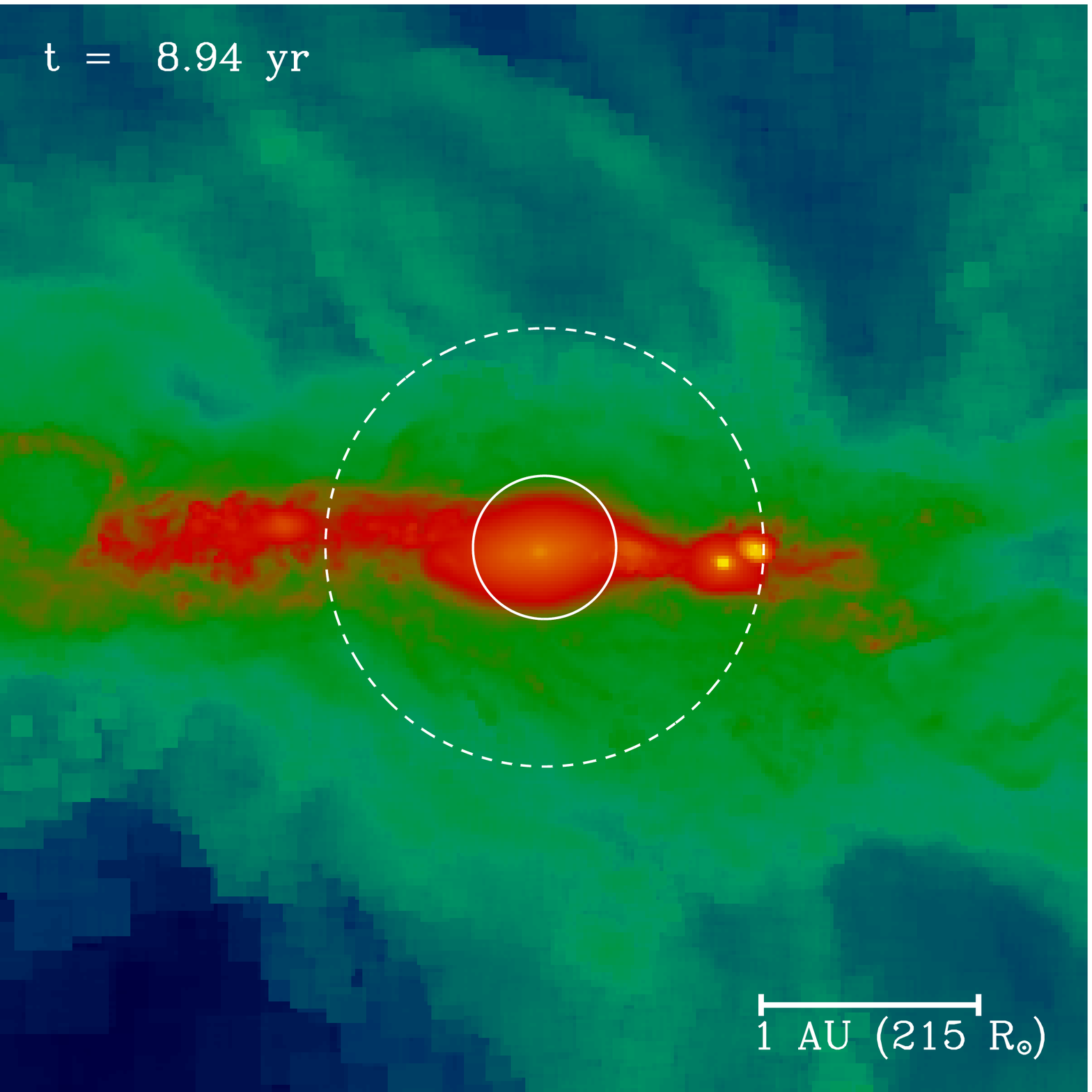}
 \caption{Same as previous figure, except with images oriented to show the disk edge-on.  Each panel has a width of 5 AU.
Note the flattened structure of the protostars at these early times.
}
\label{star_morphb}
\end{figure*}

\subsection{Convective Instability}

In Figure \ref{conv} 
we estimate the regions of the protostar that are unstable to convection, determined by the relation between $\nabla_{\rm phys}$ and $\nabla_{\rm ad}$.  As stated by the Schwarzschild criterion, convection 
in non-rotating stars
is expected if 
$\nabla_{\rm phys} > \nabla_{\rm ad}$, where

\begin{equation}
\nabla_{\rm phys} = \left( \frac{d \, {\rm ln} \, T}{ d \, {\rm ln} \, P} \right)_{\rm phys} 
\end{equation}

\noindent is the actual temperature gradient within the protostar, which we estimate directly from the simulation output.  
Furthermore, 

\begin{equation}
\nabla_{\rm ad} =  \left( \frac{d \, {\rm ln} \, T}{ d \, {\rm ln} \, P} \right)_{\rm ad} 
\end{equation}

\noindent is the adiabatic variation of temperature with pressure.  

To account for rotation and the restoring effect of angular momentum, we must instead use the related Solberg-Hoiland criterion for convective instability (e.g. \citealt{kippenhahn&weigert1990}):
\begin{equation}
\nabla_{\rm phys} > \nabla_{\rm ad} + \nabla_{\Omega}{\rm sin}\vartheta  \mbox{\ .}
\end{equation}

\noindent Here

\begin{equation}
\nabla_{\Omega} = \frac{H_{\rm p}}{g_{\rm grav} \delta} \frac{1}{\omega^3} \frac{{\rm d}\left(\Omega^2 \omega^4 \right) }{ {\rm d}\omega} \mbox{,}
\end{equation}

\noindent  where $H_{\rm P}$ is the pressure scale height, $g_{\rm grav}$ the gravitational acceleration, $\vartheta$ the colatitude angle from the axis of rotation, $\omega = r {\rm sin} \vartheta$ the distance to the rotation axis, and 

\begin{equation}
\delta =  - \left( \frac{d \, {\rm ln} \, \rho}{ d \, {\rm ln} \, T} \right)_{\rm P} \mbox{.}
\end{equation}

\noindent For solid-body rotation 
which very roughly describes our protostars within $R_*$,  this expression for $\nabla_{\Omega}$ can be simplified to

\begin{equation}
\nabla_{\Omega} = 4 \frac{\Omega^2}{g_{\rm grav}} \frac{H_{\rm p}}{ \delta }    \mbox{.}
\end{equation}

\noindent We approximate that $H_{\rm p} \sim P/(g_{\rm grav}\rho)$, where $P$ is the gas pressure \noindent (see, e.g., \citealt{maederetal2008}).
For simplicity, in our estimates we also set ${\rm sin}\vartheta$ equal to one.

If all energy transport is through radiation, we may also write $L_* = L_r$ and $\nabla_{\rm phys} = \nabla_{\rm rad}$, where

\begin{equation}
\nabla_{\rm rad} = \left( \frac{d \, {\rm ln} \, T}{ d \, {\rm ln} \, P} \right)_{\rm rad} = \frac{3}{16 \pi a c G} \frac{P \kappa}{T^4} \frac{L_{\rm r}}{M_{\rm enc}} \mbox{.}
\end{equation}

\noindent  
To account for the effects of rotation, we may furthermore replace $M_{\rm enc}$ with

\begin{equation}
M_r = M_{\rm enc} \left(1 -  \frac{\Omega^2}{2\pi G \rho_0 } \right)
\end{equation}

\noindent (\citealt{zeipel1924}) where $\rho_0$ is the average stellar density and is typically $\sim 10^{-5}$ g cm$^{-3}$ within $R_*$.  For our protostars the parenthetical factor in the above equation is of order unity and ranges from 0.8 to 0.9.

Figure \ref{conv} shows $\nabla_{\rm phys}$, $\nabla_{\rm ad}$, and $\nabla_{\Omega}$ at the end of each simulation.  
Note that the estimate of $\nabla_{\Omega}$ begins to break down outside of $R_*$, where $\Omega$ starts to deviate from a solid-body profile, so we only show  $\nabla_{\Omega}$ out to this radius.  
Throughout most of the protostar, $\nabla_{\rm ad}$ is the largest term and prevents convection within the protostars.   
$\nabla_{\rm phys}$ and $\nabla_{\Omega}$ instead maintain values closer to zero
with the exception of regions within $\sim$ 10 $R_{\odot}$. 
We emphasize, however, that improved simulations, including the proper treatment of diffusive transport of radiation, will be necessary to accurately model convection within Pop~III protostars. 
Bearing this caveat in mind, at the end of the simulations only one protostar, that of Minihalo 2, exhibits possible central convection.
Though not shown in Figure \ref{conv}, at earlier times other protostars also have short periods where the convection criterion is satisfied, giving   some indication that Pop III protostars may have typically undergone periods of convection in their cores as they grew in mass.  
However, overall the protostars appear non-convective.

Prototars on the Hayashi track are expected to have convective zones, but if the protostar is growing then the presence of an accretion shock will return it to stability against convection 
(e.g. \nocite{stahler1988b} Stahler 1988b, see also \citealt{wuchterl&klessen2001,wuchterl&tscharnuter2003}).  
However, previous one-dimensional modeling of Pop III evolution under rapid accretion (e.g. \citealt{stahler&palla1986,omukai&palla2003,hosokawaetal2010}) indicate that growing protostars can still become convectively unstable after  the onset of nuclear burning.  In particular, \cite{hosokawaetal2010} find that metal-free protostars of mass $\la$ 1 M$_{\odot}$ accreting through a disk may briefly have a central convective zone after deuterium burning begins.  However, this zone soon disappears while a longer-lasting surface convection zone emerges as the protostar grows and the D-burning region moves to the outer layers.  For spherical accretion, however, deuterium burning and convective zones do not appear until the protostars have attained higher masses ($\la$ 10 M$_{\odot}$, \citealt{omukai&palla2003}).  


Without modeling of this nuclear ignition or of radiative transfer, however, we do not expect sustained convection to appear in our simulations.  
We furthermore note that, despite the differences in opacity between Pop III, low-metallicity, and solar-metallicity protostars, appearance of convection zones associated with the onset of deuterium burning is expected regardless of metallicity.  
In all cases the main contributions to opacity are not from metals but from electron scattering and free-free absorption in stellar interiors hot enough to undergo deuterium burning, or from  H$^-$ absorption and photoionization depending on the gas temperature in cooler outer regions of the star (e.g. \nocite{stahler1988a} Stahler 1988a; \citealt{durisenetal1989, omukai&palla2003}).  However, the extra contribution to the electron fraction from metals does still lead to higher opacities in solar-metallicity gas as well as relatively smaller radii in Pop III stars (e.g. \citealt{alexanderetal1983, tumlinson&shull2000}).




\subsection{Time Evolution}

Figures \ref{star_vs_t} and  \ref{vrot_vs_t} show how various properties of the protostars evolve over time.  Mass $M_*$ and radii $R_{\rm p}$ and $R_*$ of the protostars are determined as described in \cite{greifetal2012} and Section 2.4.  In Figure \ref{vrot_vs_t}
we determine $v_{\rm rot,*}$ similarly to the method described in Equation 4, except that the sum is performed over all hydrodynamic elements within $R_{\rm p}$, and $M$ is replaced with $M_*$.
When compared with $v_{\rm Kep,*} = \left( G M_*/R_{\rm p} \right)^{1/2}$, we see that each of the stars maintains a rotational velocity that is a significant fraction of its break-up speed.  The protostar of Minihalo 1 (panel {\it a} of Figure \ref{vrot_vs_t}) even appears super-Keplerian for a brief period around 4 yr.  However, this is an artifact of our spherical averaging scheme, and the super-Keplerian period coincides with when the protostar is undergoing its first merger and exhibits a shape that is more flattened and bar-like than spherical.  
We here briefly mention that many of the secondary protostars in each minihalo also maintain very large rotation rates.  The majority of secondary protostars have values of $v_{\rm rot,*}/v_{\rm Kep,*}$ ranging from $\sim$~70-90\% at the final simulation outputs, though some have lower values of $\sim$~20-30\%.

We also quantify the level of rotational support with the parameter $\beta$, which we define as the ratio of the specific angular momentum of the protostar, $j_*$, to its Keplerian angular momentum 
$j_{\rm Kep}$.  We thus have

\begin{equation}
\beta =  j_*/j_{\rm Kep} \mbox{,}
\end{equation}

\noindent where

\begin{equation}
j_* = \frac {\left[\left(\sum m_i j_{x,i}\right)^2 +  \left(\sum m_i j_{y,i}\right)^2 +  \left(\sum m_i j_{z,i}\right)^2\right]^{1/2} } {M_*}  \mbox{.}
\end{equation}

\noindent The index $i$ refers to values for individual gas elements within the protostar, and

\begin{equation}
j_{\rm Kep} = \left( G M_* R_{\rm p}\right)^{1/2} \mbox{.}
\end{equation}

\noindent In Figure \ref{beta_vs_t}, it can be seen that each of the stars remains substantially rotationally supported and maintains a relatively constant value of $\beta \sim 30\%$, with some fluctuation above and below this value.  
The variations in $\beta$ also tend to follow the fluctuating rotational support in the surrounding disk, out to 10$R_{\rm p}$ (dotted line in Figure \ref{beta_vs_t}), with the exception of the Minihalo 1 protostar.    
In addition, the $\beta$ oscillations roughly follow the fluctuations in the ratio of rotational to Keplerian velocity at the protostellar surface $R_{\rm p}$ (dashed line in Figure \ref{beta_vs_t}).  Each protostar generally maintains $v_{\rm rot}$ of $\ga$ 80\% of $v_{\rm Kep}$ at its surface, demonstrating that the rapid protostellar rotation is set by the nearly Keplerian rotational speeds at the boundary between the inner disk and protostellar surface.  The rapid rotation at the inner disk boundary in turn originates from the angular momentum of the larger-scale disk.  This Keplerian inner disk is likely to persist to later times (see, e.g., \citealt{stacyetal2011}).  Though these simulations only follow the initial stages of protostellar accretion, as the protostars grow by an order of magnitude their high spins may be maintained through continued accretion from a nearly Keplerian disk.

This significant rotation causes the protostars to have flattened structures, as is visible in Figures \ref{star_morph} and \ref{star_morphb}, where we depict the density structure around the main protostars at the end of each simulation.  
Note that the flattening is much more significant on scales of $R_{\rm p}$ than on the smaller scales of $R_*$.
Furthermore, the main protostars of Minihalo 2 and 4 exhibit even more asymmetry within $R_{\rm p}$ than those of Minihalo 1 and 3.  
This is because in the former case, the protostars are undergoing mergers at the times shown, leading to greater flattening.
As discussed in \cite{linetal2011}, such deviations from spherical symmetry in stars undergoing disk accretion are to be expected as angular momentum is exchanged between the star and disk through gravitational torques.  
Similar to the stellar rotational evolution seen in their calculations, each protostar rapidly spins up over the first few Keplerian rotational periods $\tau_{\rm Kep}$, where in our case typically $\tau_{\rm Kep} \sim$ 1 yr.  
\cite{linetal2011} furthermore find a longer-term stabilization of the spin evolution, where the stellar spin may remain at $\sim$ 50\% of the break-up value, or may undergo a slow decline, depending on the properties of the surrounding disk.  
Their simulations address only non-fragmenting disks.
Nevertheless, even given the fragmentation and protostellar merging in our simulations, we still see a similar lack of significant evolution in $\beta$ for any of the protostars.

If the very nearby secondary protostars do not rapidly merge with the main protostars, then they may also affect the longer-term evolution of the main protostars through tidal interaction and exchange of mass and angular momentum.
For instance, if the binary components undergo a mass transfer through an accretion disk, the mass-gainer in the binary will be spun-up while the mass-donor spins down (see discussion and references in, e.g., \citealt{langer2012}).  However, some of the rotational increase of the mass-gainer will be lost due to tidal interactions between mass-transfer events, which tend to spin down the mass-gainer and to synchronize it with the orbital motion (e.g. \citealt{zahn1977}; Petrovic et al. 2005b; \citealt{langeretal2003}). This temporary spin down from tidal torques is visible, for instance, between 7 and 8 yr in panel {\it b} of Figure \ref{vrot_vs_t}. However, the overall substantial rotation rate of the mass gainer is maintained through disk accretion, and this may subsequently alter its stellar evolution as previously discussed.  

The close binary will additionally affect the angular momentum evolution of the circumstellar disk.  In particular, the action of tidal torques between the secondary and the disk will transfer angular momentum from the disk into orbital motion, thereby enabling mass accretion onto the primary (e.g. \citealt{pap&pringle1977}).  If the binary system remains wide enough, then the inflowing matter will quickly arrange into a circumbinary disk, as is already apparent in Figure \ref{star_morph}.  Spiral structure develops due to gravitational torques from the binary (e.g. \citealt{bate&bonnell1997}), and these disk perturbations allow for further angular momentum redistribution and accretion onto the binary.  Over time the binary will also tend towards an equal mass ratio between the two members.  This is because angular momentum conservation causes inflowing gas to fall onto the disk at some radius beyond the more massive member,  where it is more easily captured by the smaller companion.

\subsection{Dependence on Minihalo Characterstics}

\begin{figure}
\includegraphics[width=.48\textwidth]{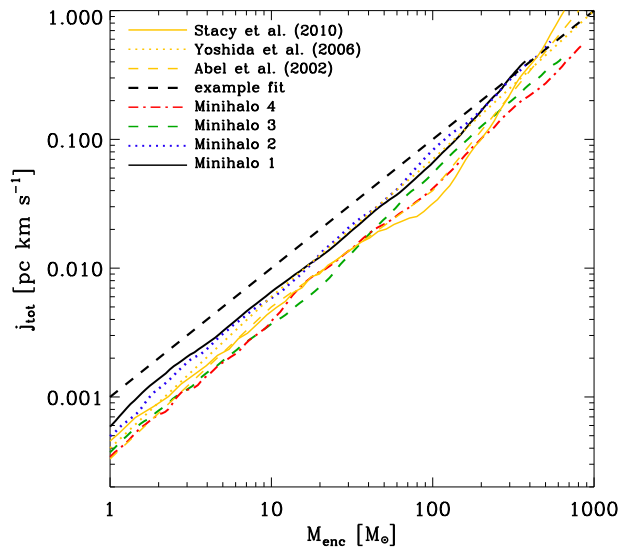}
 \caption{Angular momentum profile of gas within each minihalo.  Each minihalo is denoted by the same line style as in previous figures.  Thin yellow lines denote angular momentum profiles found in separate cosmological simulations.  Solid yellow line is taken from Stacy et al. (2010), dotted yellow line from Yoshida et al. (2006), and dashed yellow line from Abel et al. (2002).  Thick black dashed line shows an approximate powerlaw fit to these profiles, $j \propto M_{\rm enc}$. The profiles are all very similar even for a variety of cosmological realizations.}
\label{amomprof}
\end{figure}

The initial collapse of the minihalo gas from IGM to protostellar densities was discussed in detail in \cite{greifetal2012}.  They found that the radial and rotational velocity structure of each minihalo, as well as density and temperature evolution, were consistent with previous studies (e.g. \citealt{abeletal2002,brommetal2002,yoshidaetal2006,yoshidaetal2008}).  
Furthermore, the gas collapse within each minihalo generally follows the Larson-Penston solution for isothermal and self-gravitating gas (\citealt{larson1969,penston1969}).  The angular momentum profiles of the central few hundred solar masses of each minihalo are all the same to within a factor of a few, in addition to showing similar agreement to previous cosmological simulations (Figure \ref{amomprof}; \citealt{abeletal2002,yoshidaetal2006,stacyetal2010}).

\begin{figure*}
\includegraphics[width=.8\textwidth]{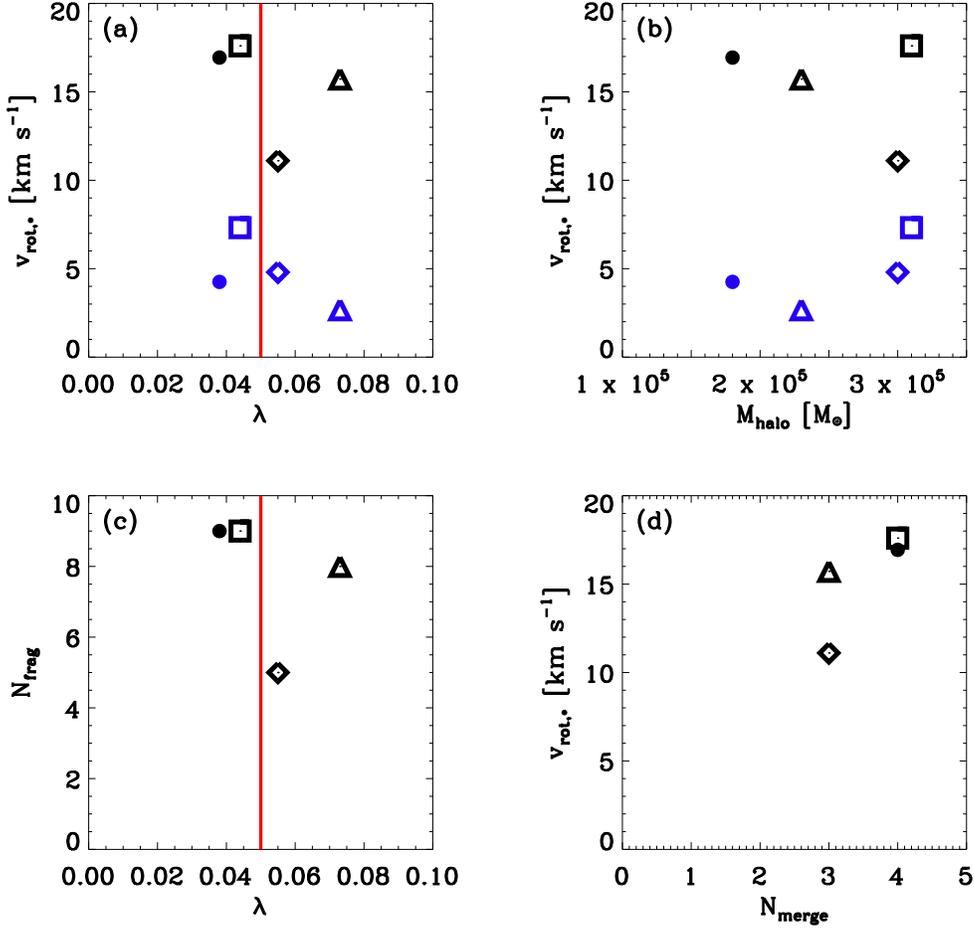}
 \caption{Comparison of the properties of the largest protostars with properties of their host haloes.
{\it (a):}  Initial and final measured rotational velocities of the largest prototar, $v_{\rm rot,*}$, versus the spin $\lambda$ of the host halo.  Initial values are in blue, while final values are in black.  Red line shows the average $\lambda$ value for DM haloes ($\lambda = 0.05$, e.g. Barnes \& Efstathiou 1987). 
{\it (b):}  Initial and final $v_{\rm rot,*}$ versus minihalo mass $M_{\rm halo}$.     
{\it (c):}  Total number of fragments $N_{\rm frag}$ formed with respect to spin of the minihalo.  
{\it (d):} Final $v_{\rm rot,*}$ versus total number of mergers with the main protostar $N_{\rm merge}$. 
The diamond represents Minihalo 1, the  triangle Minihalo 2, the square Minihalo 3, and the circle Minihalo 4.
}
\label{halo}
\end{figure*}

The minihaloes are not perfectly identical, however, and 
it is interesting to see whether variation in properties of the host minihaloes lead to any systematic variations in the properties of their protostars.  Figure \ref{halo} shows the final $v_{{\rm rot},*}$ of each star with respect to the spin parameter, $\lambda$, and mass of the minihalo.  
Also denoted is the typical spin parameter of DM haloes, $\lambda=0.05$, as measured in large-scale cosmological {\it N}-body simulations (\citealt{barnes&efstathiou1987,jang-condell&hernquist2001}),
 where $\lambda$ is defined as  
\begin{equation}
\lambda = \frac{J \left| E \right|^{1/2}}{G M_{\rm halo}^{5/2}} \mbox{\ .}
\end{equation}
\noindent Here $J$, $E$, and $M_{\rm halo}$ are the total angular momentum, energy, and mass of the halo, respectively.
We have too few sample minihaloes, however, to find an obvious correlation. 
Furthermore, when the protostars have first formed, as well as at the end of the simulations, the rotation axes of the protostars are not aligned with that of their minihaloes.  The angle between the rotation axes varies from $30^{\circ}$ to $160^{\circ}$.  
This further demonstrates a lack of connection between the spins of the protostars and their host minihaloes.
We additionally checked for relationships between the total number of protostellar fragments formed, $N_{\rm frag}$, and $\lambda$ and  $v_{\rm rot,*}$. Again, we did not simulate a sufficient number of minihaloes to find statistically significant correlations, though such an investigation with an increased number of minihaloes would be worthwhile for future work.

\section{Caveats}

We point out the caveat that magnetic fields, which were not included in these computations, may affect the angular momentum build-up of the protostellar disk and the subsequent protostellar rotation rate 
(e.g. \citealt{tan&blackman2004,silk&langer2006, maki&susa2007,schleicheretal2010, federrathetal2011,suretal2012}; \nocite{schoberetal2012a}Schober et al. 2012a, \nocite{schoberetal2012b}2012b).  
In addition, magnetic fields may  `disk-lock' the star to a certain rotation rate that depends upon its mass, accretion rate, magnetic field, and radius (\citealt{koenigl1991}; see also \citealt{shuetal1994,matt&pudritz2005}).
This may help in estimating the longer-term angular momentum evolution of the central object.  Depending on the details of such disk-locking, the large angular momentum of the disk may allow high stellar rotational velocities to be maintained beyond the phases we are modeling here, provided angular momentum is not later lost to outflows, etc.

Magnetic fields may also allow for the operation of the Spruit-Taylor dynamo within the protostar (\citealt{spruit2002}).  If this mechanism is active in Pop III stars, it can facilitate angular momentum transport from the core to the envelope.  This allows the star to maintain solid-body rotation, spinning up its outer layers and allowing the surface to reach critical rotation speed earlier in its evolution as compared to models which do not include magnetic torques.  This helps to explain the difference in the results found by, e.g., \nocite{ekstrometal2008}Ekstr{\"o}m et al. (2008) and \nocite{yoonetal2012}Yoon et al. (2012). For stars that undergo SN deaths, magnetic torques may additionally lead to a rotation rate of the collapsing iron core that is over an order of magnitude smaller than those without magnetic fields (\citealt{hegeretal2005}).
To model the formation of Pop III stars with further improved accuracy, future simulations should include the effects of magnetic fields (e.g. \citealt{turketal2012}).

Another important caveat to bear in mind is the short timescales followed in the simulation.  The simulations followed $\sim$ 10 yr, which is a small fraction of the Kelvin-Helmholtz time, $t_{\rm KH}\sim 10^5$ yr for massive stars, the typical timescale for stellar assembly. The rotational structure of a protostar only years after it has newly formed will not be the same as that when the protostar enters or later leaves the MS.  
Simulations that cover longer timescales will be required to see how the stellar spin as well as the inner accretion disk will evolve.  
Only if mass infow continues to maintain high rotational support in the inner disk, and if the inner disk continues to transfer angular momentum onto the star, can large stellar spin rates be maintained.

\section{Discussion and Conclusion}

We use the cosmological simulations of \cite{greifetal2012}, which resolve four separate minihaloes down to sub-stellar scales (0.05 R$_{\odot}$), to analyze the rotation and internal structure of Pop III protostars.  This is the first simulation to provide a direct view of Pop III protostellar structure from cosmological initial conditions.  We find the protostars quickly develop a roughly solid-body rotation profile, while their surface rotation velocities range from $\sim 80-100$\,\% of 
$v_{\rm Kep}$ at the end of each simulation.  Each of the four protostars examined maintains high rotation velocities even after undergoing multiple merger events.

The protostars generally seem non-convective over the timescales simulated. However, future calculations which include a proper prescription for radiative diffusion, as opposed to the simplified escape probablity formalism employed in Greif et al. (2012), will be necessary to properly model convection in Pop~III protostars.  
There is also little evidence of correlation between the properties of each host minihalo and the spin of its largest protostar or the total number of protostars formed in the minihalo.
More minihaloes would be necessary to derive meaningful statistics, though it would be very informative for future simulations to more thoroughly examine correlations between minihalo properties and the Pop III systems they host.

We furthermore note that future work will follow the protostellar evolution for significantly longer timescales.  This will allow for a more direct determination of how the protostars evolve, and how this evolution is affected by rotation.  
As the protostars grow in mass and continue on to the MS, rotation could alter the protostar's life in a number of ways.  
Rotation rates which persist at sufficiently high speeds, for instance, may allow for mass loss through stellar winds generated at the `$\Omega\Gamma$' limit (see discussion in, e.g., \citealt{maeder&meynet2012}), even though mass loss through line-driven winds is expected to be minimal (e.g. \citealt{kudritzki2002}).  This would reduce the final mass of the star and thus may alter the stellar death it will undergo.  

As described above, under the high stellar rotation rates we have inferred, the metal production during the lifetime of the star would also be generally enhanced 
(e.g. \nocite{ekstrometal2008}Ekstr{\"o}m et al. 2008, \citealt{yoonetal2012}).
The temperature and luminosity of the star will also be altered, and possibly greatly enhanced if rotational mixing is sufficient for the star to undergo chemically homogeneous evolution (CHE; \citealt{yoonetal2012}), though not all studies find that CHE can take place in rotating Pop~III stars (e.g.\nocite{ekstrometal2008} Ekstr{\"o}m et al. 2008).
CHE may furthermore provide a mechanism for a Pop III star to become a WR star and eventually a GRB without being in a tight binary (\citealt{yoon&langer2005,woosley&heger2006}; \citealt{yoonetal2012}).
We finally note that CHE may also lower the minimum mass at which a star will undergo a PISN death from 140 M$_{\odot}$ to$\sim$ $64$ M$_{\odot}$ (\citealt{chatz&wheeler2012,yoonetal2012})

The high spins seen in these protostars for a range of minihalo characteristics already demonstrate that a Pop III star of a given mass will have a number of possible evolutionary pathways depending upon its rotation rate.  Computational power is progressing to the point where three-dimensional simulations can begin to examine not only the mass growth but also the spin evolution of Pop~III stars. 
In complement to numerical studies,
observations of extremely metal-poor stars in the Galactic halo are key to provide constraints on the mass distribution and rotation rates of Pop III stars, their immediate progenitors (e.g. \citealt{beers&christlieb2005,frebeletal2005,caffauetal2011}). 
Much of the best evidence will come from analyzing the abundance patterns of the oldest stars found in our Milky Way or Local Group satellites, including stars within the Galactic bulge (e.g. \citealt{chiappinietal2011}).  
Further important constraints on the nature of primordial stellar populations may come from abundance analysis of damped Lyman-$\alpha$ systems (e.g. \citealt{chiappini2006,hirschi2007}), while any putative low-mass Pop~III stars that were ejected from their minihaloes may also someday be directly detected. 
These continually improving numerical and observational efforts will
allow us to probe the early Universe in ever-greater detail.

\section*{Acknowledgments}
The authors wish to thank John Mather and Sally Heap for insightful discussions.
AS is grateful for support from the JWST Postdoctoral Fellowship through the NASA Postdoctoral Program (NPP).
VB acknowledges support from
NASA through Astrophysics Theory and Fundamental
Physics Program grant NNX09AJ33G and from NSF through
grant AST-1009928.
We acknowledge the contribution
of Paul Clark, Simon Glover, Rowan Smith, Volker Springel, and Naoki Yoshida in carrying out the simulations on which our analysis is based.
Resources supporting this work were provided by the NASA High-End Computing (HEC) Program through the NASA Advanced Supercomputing (NAS) Division at Ames Research Center.

\bibliographystyle{mn2e}

\bibliography{popIIIrot}{}

\label{lastpage}

\end{document}